\documentclass[12pt]{article}
\pdfoutput=1
\usepackage[utf8]{inputenc}
\usepackage[T1]{fontenc}
\usepackage{float}
\usepackage{authblk}
\usepackage[normalem]{ulem}
\usepackage{amsmath,amssymb}
\usepackage{amsthm}
\usepackage{adjustbox}
\usepackage{graphicx}
\usepackage{booktabs}
\usepackage{mathtools}
\usepackage{fullpage}
\usepackage[mathlines]{lineno}
\usepackage{subcaption}
\usepackage[style=numeric-comp,sorting=none,maxnames=99]{biblatex} 
\usepackage[table]{xcolor}
\usepackage{verbatim}

\usepackage[colorlinks = true,
            linkcolor = teal,
            urlcolor  = teal,
            citecolor = teal]{hyperref}

\title{Characterizing collective physical distancing\\in the U.S.~during the first nine months\\of the COVID-19 pandemic}

\author[1]{Brennan~Klein\thanks{Equal contribution.\newline \indent\hspace{0.09cm}$^\S$Correspondence: \href{mailto:m.chinazzi@northeastern.edu}{m.chinazzi@northeastern.edu}}}
\author[1]{Timothy~LaRock$^*$}
\author[1]{Stefan~McCabe$^*$}
\author[1]{Leo~Torres$^*$}
\author[1]{Lisa~Friedland$^*$}
\author[1]{Maciej~Kos$^*$}
\author[4]{Filippo~Privitera}
\author[4]{Brennan~Lake}
\author[5]{Moritz~U.G.~Kraemer}
\author[6,7]{John~S.~Brownstein}
\author[8]{Richard~Gonzalez}
\author[1]{David~Lazer}
\author[1]{Tina~Eliassi-Rad}
\author[1,9,10]{Samuel~V.~Scarpino}
\author[1,3]{Alessandro~Vespignani}
\author[1,2]{Matteo~Chinazzi$^\S$}

\affil[1]{Network Science Institute, Northeastern University, Boston, MA, USA}
\affil[2]{The Roux Institute, Northeastern University, Portland, ME, USA}
\affil[3]{ISI Foundation, Turin, Italy}
\affil[4]{Cuebiq Inc., New York, NY, USA}
\affil[5]{University of Oxford, Oxford, UK}
\affil[6]{Boston Children's Hospital, Boston, MA, USA}
\affil[7]{Harvard Medical School, Boston, MA, USA}
\affil[8]{University of Michigan, Ann Arbor, MI, USA}
\affil[9]{Vermont Complex Systems Center, University of Vermont, Burlington, VT, USA}
\affil[10]{Santa Fe Institute, Santa Fe, NM, USA}

\begin{document}
\maketitle

\begin{abstract}
The COVID-19 pandemic offers an unprecedented natural experiment providing insights into the emergence of collective behavioral changes of both exogenous (government mandated) and endogenous (spontaneous reaction to infection risks) origin. Here, we characterize collective physical distancing---mobility reductions, minimization of contacts, shortening of contact duration---in response to the COVID-19 pandemic in the pre-vaccine era by analyzing de-identified, privacy-preserving location data for a panel of over 5.5 million anonymized, opted-in U.S.~devices. We define five indicators of users’ mobility and proximity to investigate how the emerging collective behavior deviates from the typical pre-pandemic patterns during the first nine months of the COVID-19 pandemic. We analyze both the dramatic changes due to the government mandated mitigation policies and the more spontaneous societal adaptation into a new (physically distanced) normal in the fall 2020. The indicators defined here allow the quantification of behavior changes across the rural/urban divide, and highlight the statistical association of mobility and proximity indicators with metrics characterizing the pandemic social and public health impact such as unemployment and deaths. This study provides a framework to study massive social distancing phenomena with potential uses in analyzing and monitoring the effects of pandemic mitigation plans at the national and international level.
\end{abstract}

\section{Introduction}
The near-ubiquity of mobile phone usage---coupled with state-of-the-art techniques for data anonymization and user privacy \cite{Oliver2020, DeMontjoye2018}---has led to unprecedented opportunities to gain insight into the social response to the COVID-19 pandemic \cite{Buckee2020, pepe_covid-19_2020, EstebanReport, nyt_cuebiq1, nyt_cuebiq2,reuters_cuebiq, google_mobility, Gao2020, Chang2021}. These types of data are useful for informing public policies and improving our understanding of human behavior by quantifying reductions in mobility and changes in consumer behavior (e.g. spending less on retail \cite{CensusRetail2020} or transitioning to a more sedentary lifestyle \cite{google_mobility}). In short, quantifying individuals' physical distancing behavior using mobile device data can give rise to an understanding of physical distancing as a large-scale \textit{collective} phenomenon---what it looks like at the macroscopic level when individuals' behaviors change at such a dramatic scale. In the case of  major crisis like the COVID-19 pandemic, the insights generated by mobility and proximity data provide researchers and policy makers with critical near real-time situational awareness that can help in managing our societal response. Furthermore, these data contribute to the debate around the effectiveness of the different policies and guidelines introduced to mitigate the spread of the disease \cite{Flaxman2020, Haug2020, DiDomenico2020, Fang2020, Perra2021, liu2021impact, hunter2021impact, brauner2021inferring}: what drives the reduction of daily interactions with others? By how much? What are the implications of the change in behavior both on the trajectory of the epidemic and on our projections of its spread?

Here, we present a framework aimed at characterizing the collective patterns of physical distancing emerging in a society through several measures of mobility and physical proximity: 1) the daily range of mobility for each user; 2) the fraction of users that commute to work; 3) the fraction of users that travel between metropolitan areas; 4) the number of unique contacts outside of home and work; and 5) the average duration of those contacts. We compute these measures over a sample of anonymized, privacy-preserving aggregated location data selected from more than 40 million mobile devices of users geolocated in the United States\ between January and September, 2020. Together, these complementary measures provide a macroscopic signature of what happens to a population when millions of individuals reduce their mobility and physical proximity. These measures allow us to provide a working definition of collective physical distancing in the United States during the first nine months of the COVID-19 pandemic, pre-vaccine era, and to quantify how it emerged---and, to some extent, persisted---following work-from-home policies, mobility restrictions, shelter-in-place orders, and other policy interventions implemented and promoted during the COVID-19 pandemic \cite{WH, nyt_stayathome, nyt_reopen}. We show that the defined measures capture relevant differences of behavior changes in urban versus rural settings, and are statistically associated with unemployment and teleworking rates. Notably, we also find that the measures characterizing reduction in individual contacts are early indicators of COVID-19 deaths. These findings suggest that the proxy measures identified here can, in turn, be used to calibrate epidemic transmission models aimed at defining the burden of the COVID-19 epidemic \cite{davis_estimating_2020, ray2020ensemble, woody2020projections, Buckee2020}. An interactive version of the results and measures included in this manuscript (as well as access to the anonymized, aggregated dataset) is made publicly available through the following online dashboard: \url{https://covid19.gleamproject.org/mobility}.

\section{Results}
In the following we considered longitudinal mobility data  provided by Cuebiq Inc., through its Data for Good program (\url{https://www.cuebiq.com/about/data-for-good/}). Cuebiq Inc.\ provides access to aggregated and privacy-enhanced mobility data for academic research and humanitarian initiatives, collected from users who have opted in to provide access to their GPS location data anonymously, through a GDPR-compliant framework (see Materials and Methods section for a more complete description of these data). 

The use of a convenience sample of mobile users to study behavioral variations in time requires caution, and we took measures to reduce two major sources of potential bias: user attrition and user selection bias. In particular, first, we measure our five collective physical distancing metrics using a stable longitudinal panel of Cuebiq Inc.\ users who were consistently active between January and June, 2020 (see Supplemental Information (SI) for details). This subset includes approximately 5.5 million anonymous users. Second, we have checked the socio-demographic representativeness of the panel and conducted two separate analyses with and without county-specific sampling weights that control for aggregated socio-demographic characteristics such as age, sex, race, educational attainment, and earnings (see SI). Our approach allowed us to create a statistically representative and stable sample of users to adequately measure collective physical distancing in the United States at national, state, and metropolitan levels of aggregation. In addition, as a robustness check, we provide the correlations between our measures of collective physical distancing with several other datasets that have been made publicly available (see SI). In this comparison, we observe the expected correlations between similarly defined measures. For example, the reduced mobility trends captured in our measures are strongly correlated with Google's stay-at-home measure as well as other comparable transit measures. Critically, the measures proposed in this work provide additional insights that do not seem to be fully captured by other publicly available datasets, especially when looking at publicly available proxies for human interactions.

\subsection{Measures of spatial mobility}
On March 16, 2020, the United States government issued guidelines promoting nonpharmaceutical interventions (NPIs) to reduce the spread of the COVID-19 \cite{WH}. Such interventions included school closures, state of emergency declarations requiring non-essential businesses to close, and shelter-in-place orders to minimize person-to-person contacts. By April 7, 95\% of people in the United States were being urged by their states' governors to stay home due to the pandemic \cite{nyt_stayathome}. To quantify the effects of these measures and the aggregated changes of human mobility across the United States we defined the following mobility indicators.

\begin{figure}[t!]
    \centering
    \includegraphics[width=1.0\columnwidth]{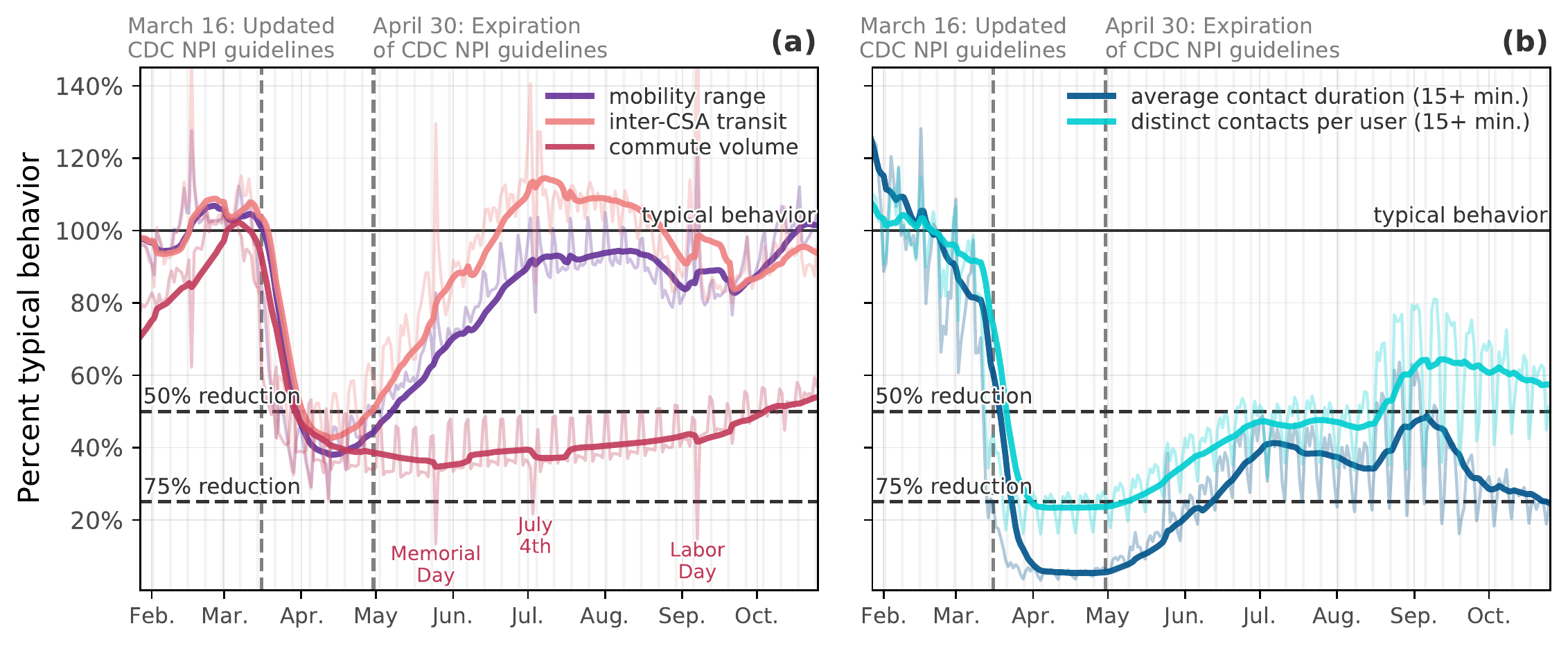}
    \caption{\textbf{Changes in mobility and person-to-person contacts over time.} Graphs show deviations from typical behavior for the same weekday in the United States. \textbf{(a)} Mobility: Individual mobility (radius of gyration), commute volume, and inter-CSA transit. \textbf{(b)} Contacts: Number of distinct contacts and average contact duration events outside of work and home. By the national declaration of emergency (March 13), reductions in spatial mobility measures had begun, reaching approximately $50\%$ of typical values by April 1; while contact measures show a reduction greater than $75\%$ by the same date. A 7-day rolling average is shown alongside each measure. Grey vertical lines denote weekends.}
    \label{fig:five_measures_collapsed}
\end{figure}

\paragraph{Short-distance traveling: daily commutes.}
We measured commuting flows by computing the total number of home to work trips originating from a given county within 24 hours. Commuting behavior is particularly interesting not only as a measure of short-range traveling but also because it can be used as a proxy for potential opportunities of transmission in the workplace. In other words, it indirectly informs us about the fraction of individuals that still go to their workplace. This information is also relevant for the modeling of disease transmission in workplace settings \cite{Aleta2020, Aleta2020_2, Mistry2020}). By early May 2020, our measure shows that across the United States there has been a reduction of approximately 65\% of the typical daily values (Figure \ref{fig:five_measures_collapsed}a, commute volume). Notably, in our data, the aggregate trend in commute volume has remained relatively stable since early May, at about a 60--70\% reduction, though it is beginning to trend upwards again as of early September. We suspect this trend is both a reflection of the reality of the ``new normal'' of work-from-home policies, along with increases in unemployment due to COVID-19 in the fall of 2020. 

\paragraph{Long-range traveling: inter-city trips.}
To study the changes in long-range traveling, we calculated the number of anonymous users who visited at least two separate Census Statistical Areas (CSAs) in a single day. In other words, we measure the volume of long-range trips between major metropolitan areas which allows us to capture---among other things---also variations in air traffic and long-range train and road trips. Indeed, by looking at inter-CSA mobility, we observe a sharp decline in the number of users traveling between CSAs (Figure \ref{fig:five_measures_collapsed}b) as compared to the baseline in every CSA included in these analysis. At its peak, the amount of inter-CSA transits among the users in our panel had decreased by almost 50\%, on average. 

\paragraph{Individual traveled distance: radius of gyration.}
Lastly, we capture the change in the range of individual daily traveled distance during the COVID-19 pandemic by calculating the \textit{radius of gyration} \cite{gonzalez_understanding_2008} for each (anonymized) mobile device in the panel of users selected for this study (see Materials \& Methods for a formal definition of the radius of gyration). This measure gives us a standardized way to tell how far an individual is traveling from their average daily position. In other words, it measures how far a user moves from their typical center of mass, most likely their home and work locations, in a given day. By early May, the average radius of gyration of users in our panel decreased by between 45--55\% relative to a typical weekday, as shown in Figure \ref{fig:five_measures_collapsed}a (mobility range). Similar results have been reported previously for New York City \cite{EstebanReport}. The range of distance traveled increases steadily throughout May and June, and by early July returns to about 95\% of the typical behavior. This increase follows the rescinding of stay-at-home orders and the steady reopening of businesses across the country, meaning increases in mobility for both employees and consumers. However, it is also likely related to increased confidence among the general public that activities requiring traveling, such as trips to the beach and hikes, could be done while practicing social distancing, making them safe to engage in. Indeed, this return to near-typical mobility range is not accompanied by a return to near-typical person-to-person contact events, giving support to the evidence that public's confidence in the safety of low-contact activities increased in time.

\subsection{Measures of contacts among individuals}
For the purpose of contact tracing, the CDC defined a \textit{close contact} as someone who was ``within 6 feet of an infected person for at least 15 minutes'' \cite{cdc_closecontact}. Using this guidance, we operationalize the definition of contact as two devices being within the same 8 digits geohash (a tile of approximately 38m $\times$ 20m) for at least 15 minutes (see Materials \& Methods) and we define two measures quantifying contact mixing between individuals. Even though defining contacts in this way can be noisy or imprecise due to the spatial resolution considered, we show that the measures introduced in this section positively correlate with key epidemiological indicators, e.g., new deaths (see Section \ref{sec:new_deaths}).

\paragraph{Number of distinct contacts outside home and work.}
As a first measure of social mixing we considered the number of \textit{distinct contacts} that a user has in a given day, outside of work or home. These contacts quantify the opportunity for disease transmission to/from distinct individuals, being at the same coffee place, interacting at a grocery store, and so on. On average, there was a dramatic decline in the number of distinct contacts that users had in a day with the onset of this decline around March 11 (see Figure \ref{fig:five_measures_collapsed}). Users in our panel had approximately 75\% fewer distinct contacts per day by mid-April. Unique contacts increased steadily starting in May and through June, leveling off for the remainder of the summer at approximately 40--50\% reduction. This trend reflects a general loosening of physical distancing, consistent with reopening of businesses as well as increased comfort with outdoor gathering. Importantly, however, we do not see a full return to typical behavior, suggesting that even faced with newly reopened amenities (shops, restaurants, etc.) people in the United States remained reluctant to return to pre-pandemic levels of social activity.

\begin{figure}[t!]
    \centering
    \includegraphics[width=1\columnwidth]{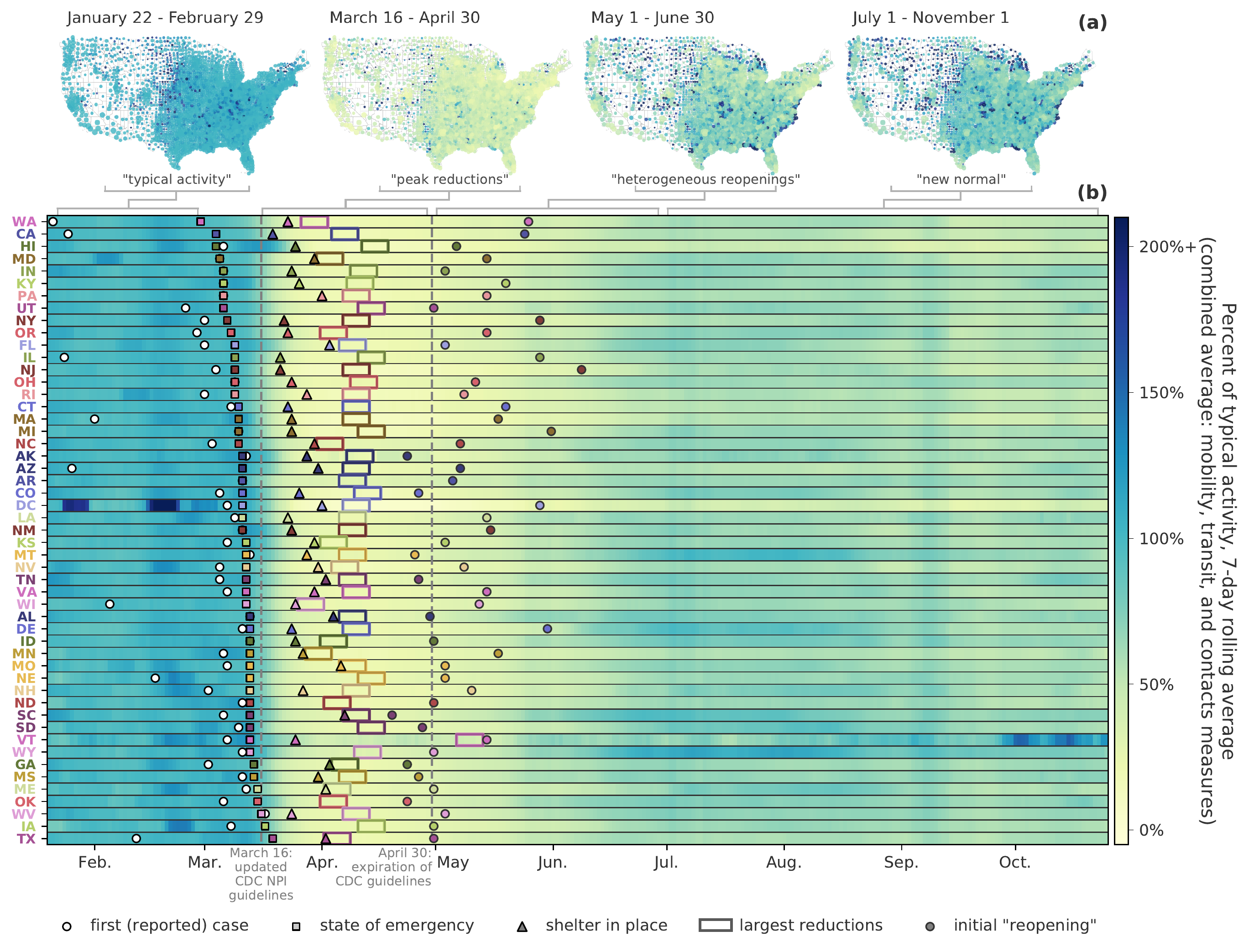}
    \caption{\textbf{The phases of collective physical distancing in the United States.} (Top) County-level maps of collective physical distancing, with each county colored by an average of its typical daily commute volume, individual mobility range, inter-CSA transit, unique contacts outside of home and work, and total duration of contacts for the time frame listed. (Bottom) Heatmap of reductions in contacts and mobility, emphasizing key dates in every state and the key phases of the pandemic in the U.S.~(reopening data from \cite{nyt_reopen}).}
    \label{fig:aggregated_reduction_states_maps}
\end{figure}

\paragraph{Average contact duration outside home and work.}
Characterizing effective contacts for disease transmission must take into account that the probability of transmission increases also with the duration of the contact. For this reason we measured each user's average total \textit{duration} of contacts with other users, based on how long their devices were located near each other. The total duration of contacts per day followed a similar pattern to the number of unique contacts. By mid-April, the duration of contacts was reduced by about 75\% compared to typical behavior before social distancing measures took effect. Through May and June there was a steady increase up to about a 45\% reduction from typical. The fact that total duration of contacts was reduced further than distinct contacts per day indicates that the increase in distinct users met is not always accompanied by an increase in time spent together. Again, some of this could be due to increased comfort with outdoor, socially distanced behavior, such as passing others on a walk through the park. Similar to the mobility range trends discussed earlier, throughout May and June there was a steady increase in contact events. However, the trend does not approach typical behavior by the end of July, instead hovering between 50--65\% of typical.

\subsection{Collective physical distancing in the contiguous United States}

The COVID-19 pandemic has brought some of the most substantial disruptions to collective human behavior in living memory. The timeline of these behavioral disruptions is clearly visible in Figure~\ref{fig:aggregated_reduction_states_maps}, where we report an aggregate index for each state computed as the average percentage reduction with respect to the typical activity of the five mobility and proximity indicators previously introduced (i.e., the expectation $\mathop{\mathbb{E}}_v$ for each of the five measures $v$ corresponding to the percent of typical activity). We can characterize four distinct phases of collective physical distancing behavior in the United States:
\begin{itemize}
    \item \emph{Typical activity}. From late January to late February, we observe a baseline period that define the typical activity
    \item \emph{Peak reductions}. During this period we observe the dramatic reduction of all indicators following the federal and state mandates and mitigations.
    \item \emph{Heterogeneous reopenings}. The time window from early May to late June was marked by states' reopening of businesses and schools according to different schedules and strictness in residual NPI's.
    \item \emph{New normal}. From July onward when mobility range and inter-CSA transit increased to values comparable to pre-pandemic levels, while commuting flows (i.e., people going to work) and contact measures generally remained at values lower than typical pre-pandemic levels, characterizing a new stage for people living through a pandemic.
\end{itemize}
During these four time frames, we see a combination of nationwide reductions in commuting volume to/from work. People's daily social routines changed dramatically as well, with daily mobility being reduced by up to 60\% in April, along with approximately 80\% fewer contacts with others per day at the peak of physical distancing. It remains a challenge to identify any single cause of these changes in behavior. However, when looked at together, they offer a way to characterize the evolution of our collective behavior, giving us a baseline for understanding how societies react to such a massive disruption. In the SI, we provide measure-specific curves for each state (Figure \ref{fig:allstates}) and for several major   metropolitan areas (Figure \ref{fig:city1} to Figure \ref{fig:city17}). 

\paragraph{Commuting, working from home, and unemployment}
Commuting volume decreased dramatically in early March, and, in most states, did not increase in the same way that the other mobility/contact measures have. This is likely due to several reasons, from the historic waves of unemployment in the spring and early summer, to a dramatic increase in teleworking. Indeed, throughout the COVID-19 pandemic in the U.S., we see a strong negative correlation between the percentage of teleworkable jobs and commute volume in metropolitan statistical areas (MSAs). For public health officials planning for future pandemics, this relationship between telework and commute volume is especially insightful; many of the NPIs that have been introduced throughout the pandemic were designed to limit the amount of workplace infections, and as such, reductions in commute volume that are due to increased telework (as opposed to increased unemployment) could illustrate an ideal balance of economic and public health priorities. 

\begin{figure}[t!]
    \centering
    \begin{subfigure}{1.0\columnwidth}
        \centering
        \includegraphics[width=1.0\columnwidth]{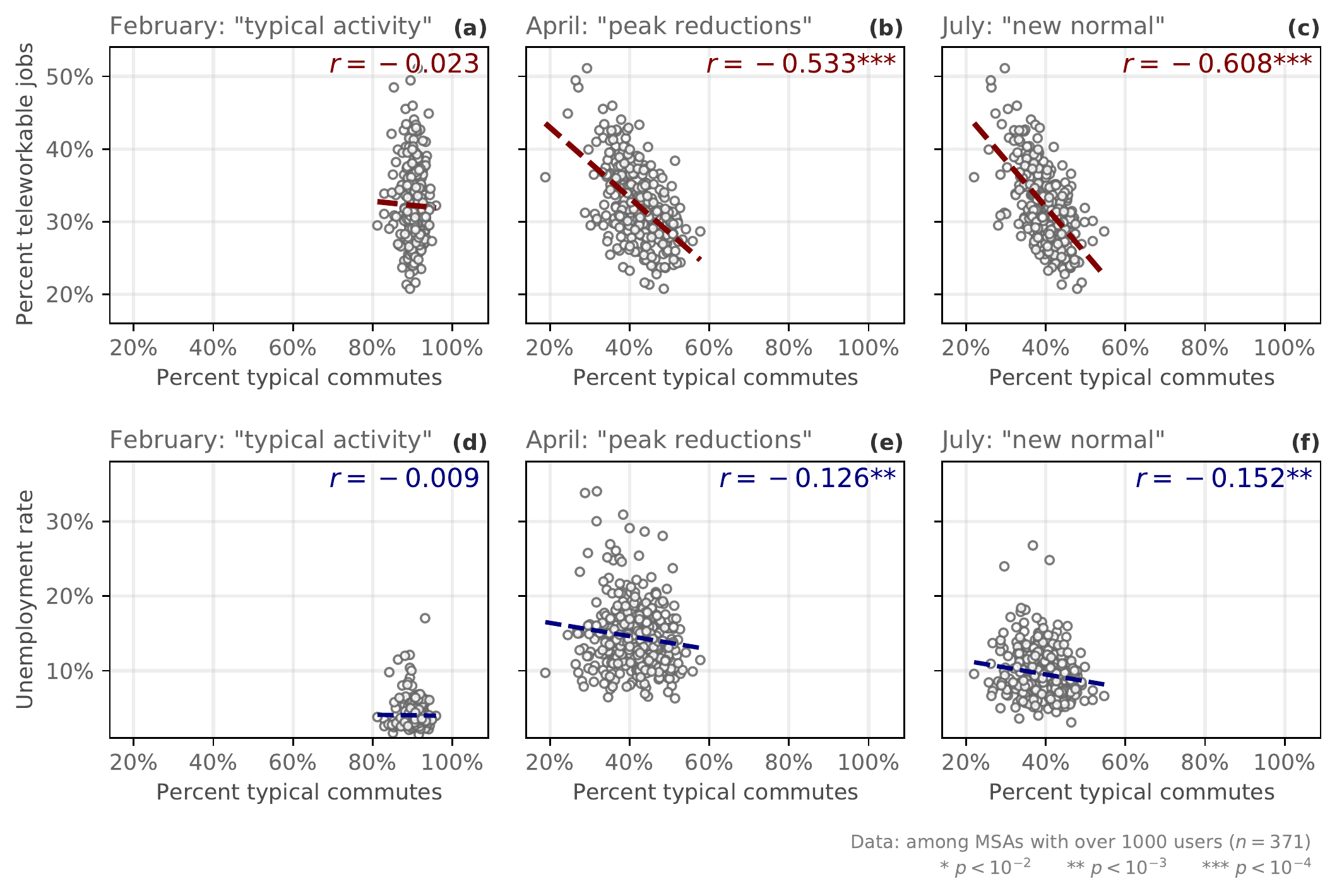}
    \end{subfigure}
    \caption{\textbf{Unemployment, teleworking and commuting patterns.} Grouping county-level employment data to the Metropolitan Statistical Area (MSA), we correlate commute volume with the percent of jobs that can readily transition to teleworking (top row) and unemployment rate (bottom row) over time. \textbf{(a \& d)}: February, during the baseline period; \textbf{(b \& e)}: April, during the peak lockdown; \textbf{(c \& f)}: July, after unemployment declined but commuting remained low---during the ``new normal'' phase).}
    \label{fig:unemployment}
\end{figure}

During the COVID-19 pandemic, many employers eliminated in-person interactions, although many jobs in the United States cannot easily transition to remote work. In 2018, the U.S.~Bureau of Labor Statistics estimated that almost 25\% of workers could work from home \cite{work_from_home_2019}, a number that varies widely by race, education level, and industry \cite{Dingel2020}. Despite widespread video-conferencing and teleworking software, it is difficult to quantify the ubiquity of these practices across the United States, though major Internet Service Providers reported traffic increases between 20--30\% \cite{nyt_internet_traffic} early on in the pandemic; various surveys have been conducted trying estimate this number as well \cite{Brynjolfsson2020}. Also, crucially, the typical commuting patterns of millions of people in the United States were impacted by an unprecedented surge in joblessness; over 42 million unemployment claims were filed in the United States between March and June \cite{weekly_jobs_apr2020}.

Our measurements allow us to explore the relationships between commuting, unemployment, and teleworking by combining our commute metric with two additional data sources: the Bureau of Labor Statistics's Local Area Unemployment Statistics (LAUS) dataset, which provides monthly estimates of county-level unemployment rates; and Dingel and Neiman's estimates of the proportion of jobs in an area that can be feasibly worked from home \cite{Dingel2020}. The Dingel and Neiman teleworkability estimates are at the Metropolitan Statistical Area (MSA) level, and the LAUS data at the county level. For this reason, in our analysis, we aggregate our measurements to the MSA level, excluding rural counties. Note that while local unemployment and commute volume vary month-to-month, the estimated proportion of teleworkable jobs is largely a static quantity.

In Figure~\ref{fig:unemployment}, we present the relationship between commute volume, unemployment, and teleworkability in February, April, and July, with each month corresponding to a distinct phase of the pandemic. In February---the baseline period for which we use to define ``typical'' activity---there is no correlation between the percent of typical commutes in MSAs and the percent of teleworkable jobs, nor is there a correlation between commuting and unemployment at that point (Figure \ref{fig:unemployment}a and \ref{fig:unemployment}d). This is expected, as the United States had not yet experienced large disruptions resulting from the COVID-19 crisis. Then, in April, following the updated guidelines about physical distancing and the massive surges in unemployment, commuting volume dropped substantially across the United States (on average, about 40\% of baseline levels, Figures \ref{fig:unemployment}b, \ref{fig:five_measures_collapsed}a). During this lockdown period, the percentage of jobs that can transition to telework show a -0.533 correlation with commute volumes (Figure \ref{fig:unemployment}b); we also observe a significant negative correlation between commute volume and unemployment rate during this period (Figure \ref{fig:unemployment}e).

\begin{figure}[t!]
    \centering
    \begin{subfigure}{1.0\columnwidth}
        \centering
        \includegraphics[width=1.0\columnwidth]{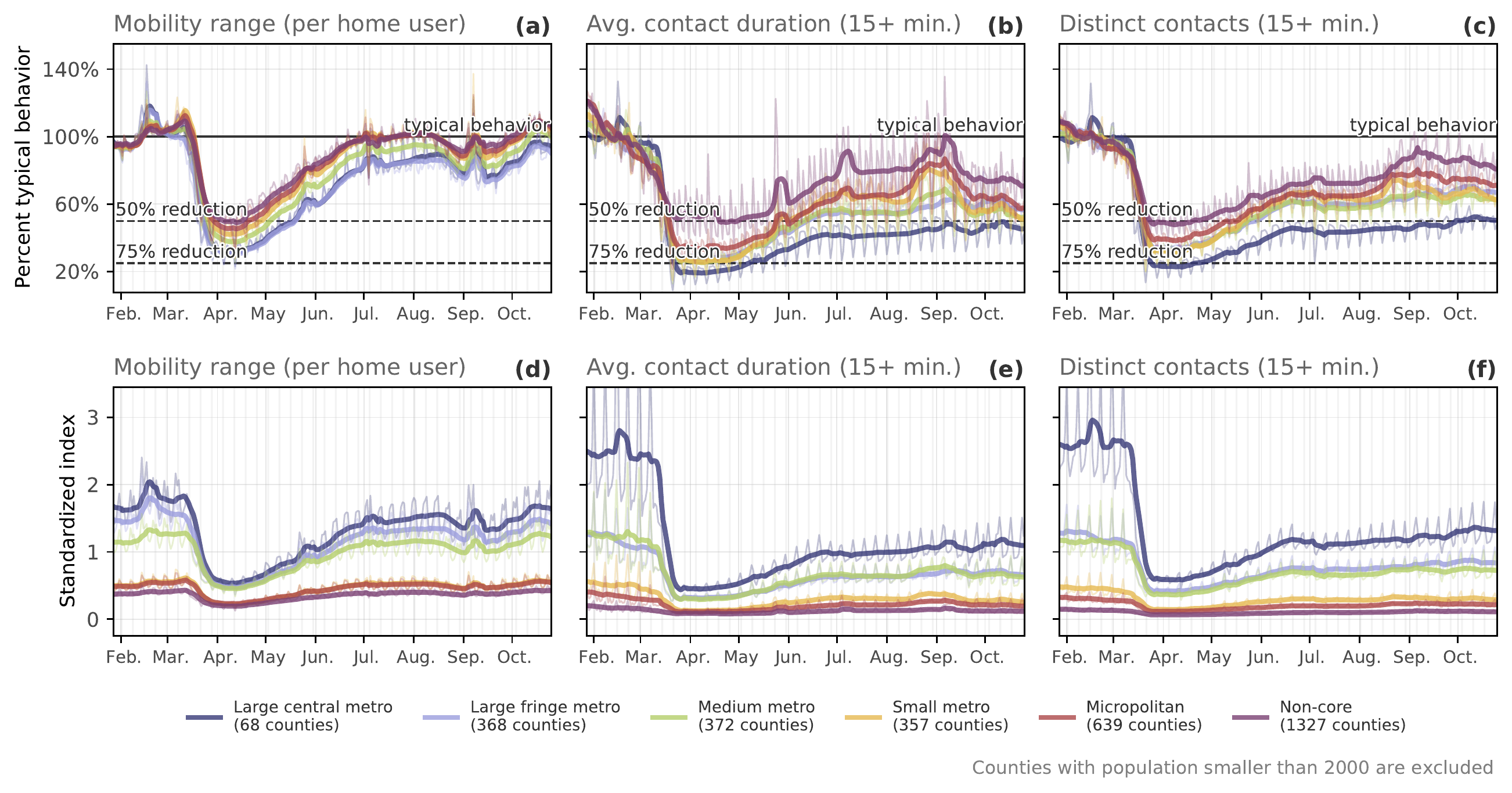}
    \end{subfigure}
    \caption{\textbf{Differences in county-level behavior based on rural-urban codes.} Each county in the United States is assigned a rural-urban code, ranging from 1 (large central metro) to 6 (highly rural, ``non-core'' counties). We average the percent of typical behavior per user (top row) and a standardized index (bottom row) across counties grouped by these six rural-urban code designations. The standardized index obscures raw values but preserves relative differences between groups; we do so by normalizing by the median value across all counties. \textbf{(a \& d)}: mobility range; \textbf{(b \& e)}: contact duration; \textbf{(c \& f)}: distinct contacts. Seven-day rolling averages are plotted in bold above raw values plotted as thin curves.}
    \label{fig:urban_rural}
\end{figure}

\paragraph{Collective physical distancing in rural and urban areas}
We also observe different levels of collective physical distancing in different parts of the country, which reflects the heterogeneity in policy response, disease incidence, geography, and population structure across the US (see e.g.~\cite{Rader2020}). By grouping our collective physical distancing measures with the National Center for Health Statistics (NCHS) urban-rural county classification scheme \cite{nchs_urban_rural}, we can compare the responses of people living in urban versus rural settings. We observe that large central metro (code 1) areas showed the largest reductions in our collective physical distancing measures for contacts and mobility range (Figure \ref{fig:urban_rural}a-c); the more rural the county, the less reduction from typical behavior. However, users living in more urban counties also had lower baselines for these measures (Figure \ref{fig:urban_rural}d-f), which we show using a standardized index as opposed to the ``percent of typical'' values shown in Figure \ref{fig:urban_rural}a-c. As a point of comparison, by the beginning of April, the median reduction of the number of distinct contacts for users in large/medium metro counties approached a level similar to the baseline of a typical user in rural, micropolitan counties (Figure \ref{fig:urban_rural}f).

More-urban areas also began physical distancing behaviors earlier than more-rural areas; for example, large central and large fringe metro areas dipped to 80\% of typical for their contact measures about five days earlier than small metropolitan areas, micropolitan areas, and non-core areas. During the first week of May, micropolitan and non-core counties showed mobility range that was around 75\% of typical, while large central and fringe metro areas remained at around 50\% of typical (Figure \ref{fig:urban_rural}a). A similar rural-urban gap is seen in the percent of typical behavior for both measures of contacts (Figure \ref{fig:urban_rural}b-c).

\paragraph{Collective physical distancing and the toll of COVID-19}\label{sec:new_deaths}
Lastly, we validate the use of the proximity/contact measures introduced in this manuscript as coarse-grained approximations for true person-to-person contacts. As such, we would expect to find a positive correlation between these measures and key epidemiological indicators, such as new deaths. More specifically, we would expect that a \textit{lagged} correlation would best capture this relationship since it accounts for: a) the time from exposure to symptom onset (about 6 days); b) the median number of days from symptom onset to death (between 13 to 17 days depending on the age group considered); and c) the median number of days from death to reporting date (varying from 19 to 21 days depending on the age group) \cite{cdc_planning}. Because of these factors, the median delay we would expect in the correlation between our proxies for contacts and new deaths is in the range $38$--$44$ days. Nationwide, we see that this is indeed the case.

\begin{figure}[t!]
    \centering
    \includegraphics[width=0.90\columnwidth]{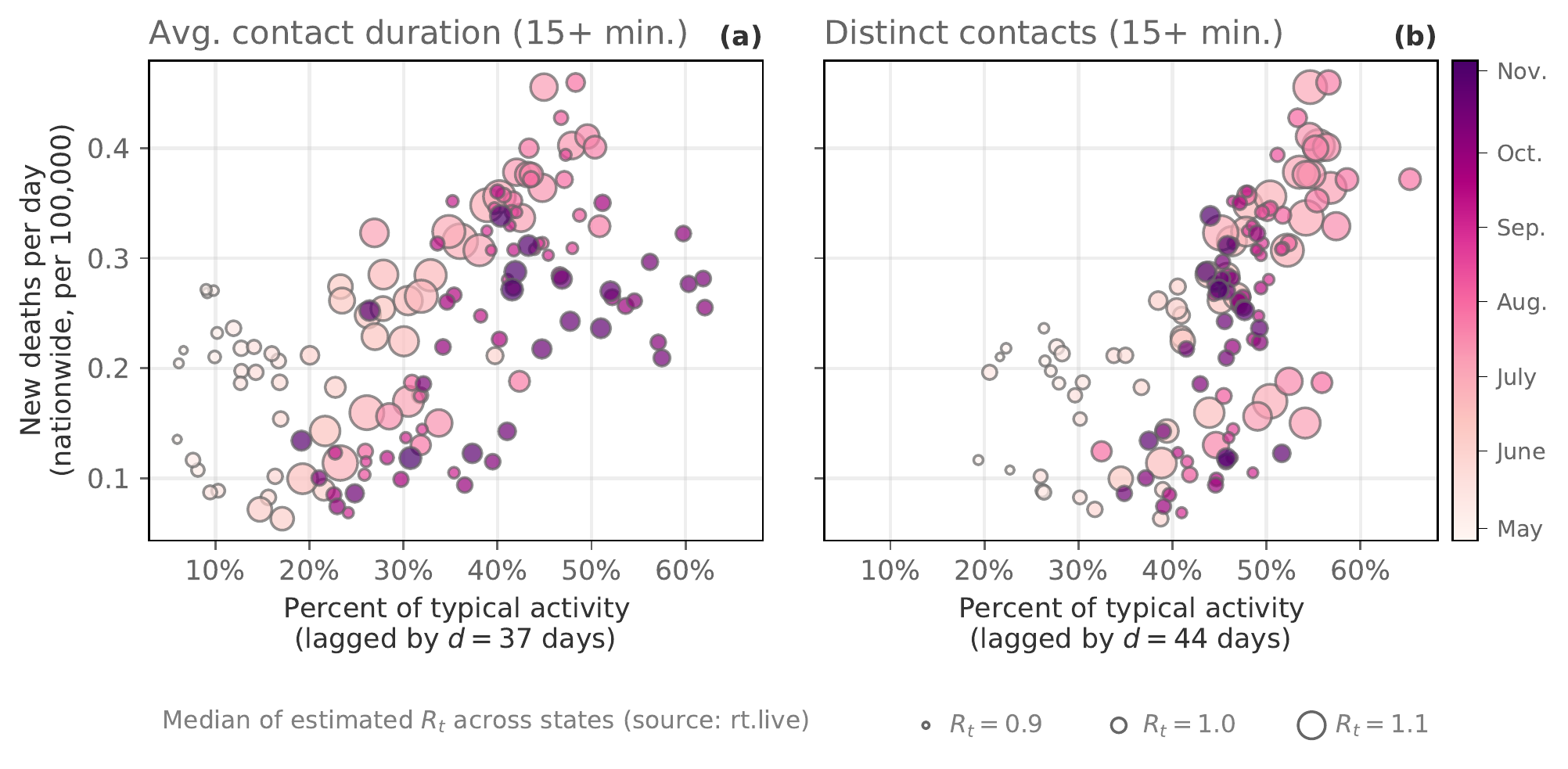}
    \caption{\textbf{Collective physical distancing and new deaths.} Here we correlate daily contact measures nationwide with new reported deaths \cite{CovidTracking} between April 30 and November 5, 2020. The horizontal axes correspond to the percent of typical contact patterns, while the vertical axis corresponds to the (lagged) number of new deaths per 100,000. \textbf{(a)} Average contact duration \textbf{(b)} Distinct contacts. A lag of $d$ days was selected for each state so as to maximize the correlation between new deaths and contact measures. Maximum correlation is observed at $d \in [37,44]$ ($d=44$ is visualized) days that is consistent with CDC estimates \cite{cdc_planning} that account for disease dynamics and reporting delays. In each subplot, darker colors indicate later dates and marker size corresponds to an estimate of the median effective reproductive number ($R_t$) across all 50 states and District of Columbia (source: \url{rt.live}). These contact measures are also positively correlated with new reported cases (but at a shorter lag, see SI Figure \ref{fig:newcases}).}
    \label{fig:newdeaths}
\end{figure}

In Figure \ref{fig:newdeaths}, we plot the nationwide percents of typical average contact duration (a) and distinct contacts (b) against the daily number of new deaths per 100,000 people nationwide. The contact measures are correlated at a delay $d \in [37, 44]$ days for each measure, which was selected by maximizing the correlation between contact patterns and new deaths. We do the same comparison for the number of new infections in the SI (Figure \ref{fig:newcases}), highlighting the robustness of this correlation nationwide.

The color of the markers in Figure \ref{fig:newdeaths} corresponds to time; darker colors indicate later dates. Here, we see an important relationship between our contact measures and the course of the pandemic. Namely, as contact patterns increased in the early summer (lighter colored markers), new infections and new deaths followed; this, in turn, was followed by \textit{decreases} in contact events, followed again by decreases in new infections by late August (darker colored markers). This is approximately the same time as when the curves in Figure \ref{fig:five_measures_collapsed}b (the contact measures) started to level off, while mobility and inter-city transit continued to rise (Figure \ref{fig:five_measures_collapsed}a). What this disconnect between mobility and contact patterns suggests is that our collective social behavior can reduce the rate of new infections and, as a result, new deaths. This finding is possibly trivial to epidemiologists and public health officials, but it is nonetheless important for our understanding of how our collective behavior impacts the trajectory of a pandemic, to validate our contacts measures as proxies for true person-to-person contacts, and it is also consistent with other findings throughout the literature on COVID-19 \cite{Xiong2020, Chang2021, Badr2020, Monod2021}. The ability to measure these patterns in almost real time shows the potential benefits of using mobile device data in forecasting (or ``nowcasting'') the trajectory of a virus, and moving forward, they present a baseline for our collective behavioral response to future pandemics.

\section{Discussion}
The massive efforts to comply with the CDC's physical distancing guidelines have come at a substantial cost to the economic and social well-being of people in the United States. By quantifying these nationwide behavioral changes, we get a glimpse into the relationship between large scale collective behavior and the course of the pandemic. Learning from these patterns is necessary to prepare for future pandemics; most notably because despite large-scale collective physical distancing, during the time window from February to December 2020, the United States has reported over 13 million cases of COVID-19 and, as of December 2020, over 340,000 reported deaths (with estimates of the true number being far higher \cite{nyt_excessdeaths}). This suggests that in the pre-vaccine era, the timing, magnitude, and synchrony \cite{Althouse2020, holtz2020} of collective physical distancing in the United States was ultimately insufficient to completely mitigate the nationwide outbreak. This is especially true when physical distancing is not combined with a vigorous testing and contact tracing regimen \cite{Aleta2020}, as was the case in countries like South Korea, Taiwan, and China.

During the ``new normal'' period from July to December 2020, there were millions of new cases and hundreds of thousands of new deaths in the United States; during this same time period, we see mobility patterns return to 100\% of baseline levels while contacts remained at around 65\% of typical activity. This suggests two key things: First, a national average of approximately 65\% of typical contacts was not sufficient for avoiding the large number of cases seen during that period. Further modeling efforts are needed to estimate the potential effects that larger decreases in contacts would have had (e.g. 60\%, or 50\%, etc. instead of 65\%). Second, this suggests that over the course of the pandemic, people may have learned to adapt their behavior in a way that allows them to travel while still limiting opportunities for contact with others. For example, visiting a park or hiking are activities that are likely associated with higher mobility but not necessarily more contacts. Indeed, in many cities across the United States, we see a relative rise in visits to parks \cite{google_mobility} during this time period. Learning from this might inform goals or benchmarks for policy responses to this or future pandemics.

In this manuscript, we quantified the unprecedented behavioral response to COVID-19 in the first 9 months of the COVID-19 pandemic in the United States---collective physical distancing at a nationwide scale---using five different measures of mobility and contact patterns. By studying the daily mobility patterns of millions of anonymous mobile phone users, we show how people altered their typical behavior, limiting daily interactions with others to comply with policy interventions and in an effort to reduce their chances of becoming infected with the virus. Understanding precisely and quantifying how individuals' behavior changed over the course of the pandemic is critical, and in this work we present several measures that transform large-scale mobile device data into near real-time epidemiological insights. Of particular importance, the contact proximity measures introduced here correlate with the onset of new deaths nationwide; this correlation is maximized at a delay of 37--44 days, in line with the range reported by the CDC \cite{cdc_planning}.

Recent work has shown that a more nuanced understanding of typical human mixing patterns can have dramatic effects on the spread of a disease and our models of the spread of a disease; it is particularly useful to understand age-based, setting-specific contact patterns within a population \cite{Mistry2020, Zhang2020}. The current study is limited by the absence of this data, and in many ways traditional surveying methods may offer more robust estimates (see \cite{Zhang2020}). However, the measures of collective physical distancing behavior that we introduce can be potentially generalized by using differences in Census tracts age distributions to estimate (on aggregate) age-specific mobility and contact reductions. Lastly, we quantify contacts based on geographic proximity and we do not attempt to link locations to information about the \textit{setting} where these contacts take place in (i.e., at a restaurant, workplace, park, etc); this information is particularly relevant because the odds of disease transmission are much higher with contacts in closed spaces compared to open-air environments \cite{nishiura2020closed, Aleta2020_2}. This can be addressed by measuring contact events within a pre-identified list of points-of-interest. 

Taken together, our lives and everyday activities have been fundamentally reshaped during the COVID-19 pandemic. Defining and quantifying the collective physical distancing that took place over the course of the COVID-19 pandemic becomes especially vital when planning for mitigation strategies in the future.

\section{Materials \& Methods}

\subsection{Description of data sources}

\paragraph{Mobile device data}
Mobility data are provided by Cuebiq Inc., a location intelligence and measurement company. Through its Data for Good program (\url{https://www.cuebiq.com/about/data-for-good/}), Cuebiq Inc.\ provides access to aggregated and privacy-enhanced mobility data for academic research and humanitarian initiatives. These first-party data are collected from users who have opted in to provide access to their GPS location data anonymously, through a GDPR-compliant framework. In order to preserve users' privacy, Cuebiq Inc.\ adds noise to users' ``personal areas'' (i.e. home and work locations) by up-leveling the coordinates of these areas to the centroid of their corresponding Census block group \cite{u.s._census_department_glossary_2019}. This allows for demographic analysis while obfuscating the true home and work location of anonymous users and preventing misuse of data.

\paragraph{Demographic and employment data} County-level demographic data, including the rural-urban designation as well as demographic data used for statistical corrections in the SI are from the United States Census and the American Community Survey (\url{https://www.census.gov}). County-level unemployment data are from the United States Bureau of Labor Statistics. Data about the percent of teleworkable jobs are from Dingel and Neiman \cite{Dingel2020}, where they estimate the percent of jobs that can transition to telework for a given Core-Based Statistical Area based on the occupation distribution within the each region.

\paragraph{State-level COVID-19 testing data and reopening data} Data about the COVID-19 testing and cases are from the COVID Tracking Project \cite{CovidTracking}, which compiles data directly from state health authorities. Data about the dates that states initially began to reopen was collected from the \textit{New York Times} \cite{nyt_reopen}.

\subsection{Collective physical distancing measures}

Below we define the different mobility and contact measures used in this work. Note also that we convert each of these measures to a per-user measure. We do this by dividing the cumulative value for each measure at each spatial resolution (e.g. a state) by either the number of users with ``home'' locations in that region (for commute volume, inter-city transit, and mobility) or by the number of users with contacts in that region (for average contact duration and number of distinct contacts). In addition, in this manuscript, we report the metrics as percent of \textit{typical} activity. We select the period between January 16 and February 28, excluding holidays, as our baseline, therefore defining what constitutes, in the context of this work, \emph{typical behavior}. Then, for each measure, we divide its daily value by the average value of its corresponding day of the week (i.e., Mondays are compared to the average Monday). In other words, values of 100\% denote typical behavior. 

\subsubsection{Estimating daily commute volume}\label{personal_areas_classification}

Cuebiq Inc.\ provides a list of obfuscated ``personal areas'' for each user. Observations geolocated from within these locations are deemed to be coming for either the home or the work location of the individual and are therefore up-leveled to preserve user privacy. That is, these coordinates are aggregated to the centroid of the Census block group level that each observation falls into. In order to quantify the changes in commuting behavior to and from work, we classify personal areas into the home or work location to be able to count commute flows. In particular, we consider the most commonly-visited personal area during nighttime hours (9:00pm -- 5:00am) as the home location of the user, while the most common non-home personal area visited during daytime hours (9:00am -- 5:00pm) is classified as the work location of the user. This method is imperfect (i.e., it may obfuscate users who exclusively work night shifts), but it is based on assumptions about the typical worker in the United States. Then, one \textit{commute} is defined as a user visiting their ``home'' and ``work'' in a given day. Lastly, in this study we take as reference the definition and location of personal areas as identified in the period immediately prior to the lockdown measures. Therefore, our commute metric reflects changes with respect to the \emph{status quo} existing prior to the COVID-19 pandemic.

\subsubsection{Estimating inter-CSA transit}

As described previously, we estimate the change in inter-CSA transit by calculating the number of anonymous users who visited at least two separate CSAs in a single day. Such inter-CSA transit could stem from long-range commutes, from travel (i.e., on federal holidays, such as Presidents' Day in February, we observe a spike in inter-CSA transit, suggesting tourism or vacation), or other miscellaneous transit including, for example, airline travel, train or bus trips, or long-range road trips.

\subsubsection{Estimating individual mobility using the radius of gyration}

As defined in \cite{gonzalez_understanding_2008}, the radius of gyration characterizes the extent of a given user's trajectory in a single day. It measures the mean square distance from the trajectory's center of mass to the locations reached that day. Formally,
\begin{equation}
    r = \sqrt{\frac{1}{n}\displaystyle\sum_{i=1}^{n} \| \vec{r}_i^{\,} - \vec{r}_{cm}^{\,} \| ^2},
\end{equation}
where $n$ is the user's number of observations on that day, $\vec{r}_i^{\,}$ is the $i^{th}$ observed position of the user, $i = 1, 2, ..., n$, and $\vec{r}_{cm}^{\,} = \sum_i \vec{r}_i^{\,} / n $ is the center of mass of the trajectory. A larger radius of gyration corresponds to a trajectory with positions that are far away from the trajectory's average position. In the current context, a smaller radius of gyration indicates that a user travels less distance away from their daily average position in a city. In order to compute typical mobility within a given region, we sum the total daily radius of gyration that by users in that region. The radius of gyration is a measure that can be interpreted as the variance around the day's center of mass and there; as with most measures defined here, there exist other measures of an individual's typical path deviation over multiple days that may provide additional insight \cite{Nielsen2022}.

\subsubsection{Estimating daily contacts outside personal areas}

The method for estimating contacts outside personal areas (i.e. outside users' home and work locations) is as follows. For each location event (aka a ``ping'') recorded, we associate its longitude-latitude coordinates to an 8-character \textit{geohash}. A geohash is a short string of letters and digits that allows to encode coordinates into a hierarchical spatial data structure that tessellates the world surface into a grid. In our case, we consider geohashes at an 8-digits resolution which encode rectangular cells of dimensions that are approximately 38m $\times$ 19m at the equator \cite{geohash_wiki}. We define two users to be co-located if they are observed in the same geohash for at least 15 consecutive minutes. Note that this duration can be set arbitrarily, but we use 15 minutes following CDC guidance \cite{cdc_closecontact}. For each user, we compute the number of \textit{unique} users that a device is in contact with during a single day and the total dwell time of their daily contacts that fit the above criteria. We average these values across users in a given region to arrive at county, state, and nationwide daily average contact duration and average number of distinct contacts.

\section*{Additional information}

\subsection*{Acknowledgements}
We thank Ciro Cattuto, Michele Tizzoni, and Zachary Cohen for their help understanding the details of Cuebiq data and Esteban Moro for his comments. We also thank Chia-Hung Yang for coding assistance. We thank Agastya Mondal and Robel Kassa for the development of the online dashboard. MC and AV acknowledge support from COVID Supplement CDC-HHS-6U01IP001137-01 and Google Cloud and Google Cloud Research Credits program to fund this project. The findings and conclusions in this study are those of the authors and do not necessarily represent the official position of the funding agencies, the National Institutes of Health or U.S.~Department of Health and Human Services. TER, LT, and TL were supported in part by NSF IIS-1741197, Combat Capabilities Development Command Army Research Laboratory under Cooperative Agreement Number W911NF-13-2-0045, and Under Secretary of Defense for Research and Engineering under Air Force Contract No.~FA8702-15- D-0001. The IRB/oversight body that provided approval for the research described here is: Northeastern University, the Office of Human Subject Research Protection (HSRP). IRB Exemption number: 20-03-23.

\subsection*{Data and code availability}
The mobility, commuting, and contact indexes introduced in this publication are accessible at the following live dashboard: \url{https://covid19.gleamproject.org/mobility}. Instructions to request the raw data for the national and state-level metrics can be requested at the previous website. Accompanying codes, as well as the county-level weights (see SI), will be made available at \url{https://github.com/mobs-lab/covid19-mobility}.

\begin{sloppypar}
\printbibliography[title={References}]
\end{sloppypar}

\clearpage

\appendix
\setcounter{figure}{0}
\setcounter{table}{0}
\setcounter{equation}{0}
\renewcommand\thefigure{\thesection.\arabic{figure}}
\renewcommand\thetable{\thesection.\arabic{table}}
\renewcommand\theequation{\thesection .\arabic{equation}}
\begin{refsection}

\section{Supplemental Information}

\subsection{Panel of Cuebiq users}

\subsubsection{Panel composition}

In this study, we analyzed the mobility and physical distancing behavior of a panel of over 5.5 million United States-based Cuebiq Inc.\ users that were active between January 7$^{th}$ and June 30$^{th}$, 2020. Specifically, users are included in the panel if all the following conditions apply. First, each user must be active for at least 21 days in each month from January until June 2020. Second, on average, each user must have at least one ping per hour (averaged over the number of days in which a user is active). Lastly, the average geolocation accuracy for each device needs to be less than 50 meters for the period of coverage.

In the values we report throughout this article, we assumed that the panel is stable over time. That is, we correct for users who might not present in the panel in the months July--October, 2020. We use the number of observed panel users every week to perform a correction to the raw values of each of the five mobility and proximity measures. To arrive at these corrected values, we simply divide each of the raw totals by the fraction of the panel size that had been seen that week (i.e. if a given county reported 100 commutes during a week where only 90\% of the panel had been active so far, then we adjust the raw number of commutes to be $\frac{100}{0.9} = 111$ commutes). Because we define ``typical'' as deviation from the total aggregated value for each measure (i.e. the typical number of commutes for a given county is defined as the sum of the number of commutes \textit{to} that county in a day), it is necessary to assume this panel stability. Finally, we weight our metrics to increase the degree of representativeness of our findings with respect to the U.S.\ population both at the national and state level aggregations. The procedure followed is described in the following section. 

\begin{figure}[t!]
    \centering
    \begin{subfigure}{0.9\columnwidth}
        \centering
        \includegraphics[width=1.0\columnwidth]{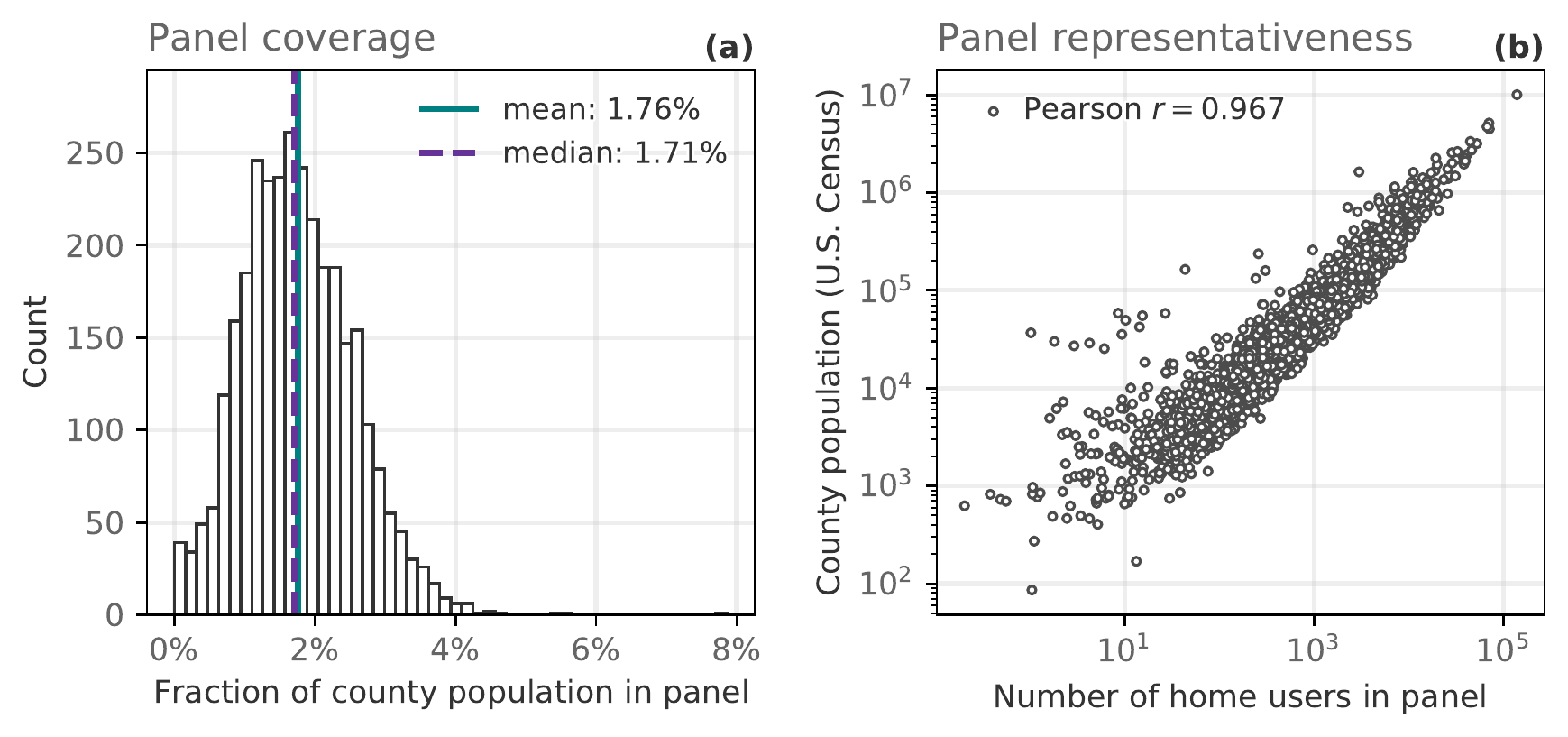}
    \end{subfigure}
    \caption{\textbf{Panel membership coverage.} By inferring counties of users’ ``home'' personal areas from the data, we can see the extent to which we are over/under- representing users on a per county basis. \textbf{(a)} Histogram of the fraction of population included in our panel of users for each county. \textbf{(b)} Scatterplot correlating the number of home users in a county against the total population of the county.}
    \label{fig:county_coverage}
\end{figure}

\subsubsection{Panel statistical representativeness}\label{sec:si_reweight}

\begin{figure}[ht!]
    \centering
    \begin{subfigure}{1.0\columnwidth}
        \centering
        \includegraphics[width=1.0\columnwidth]{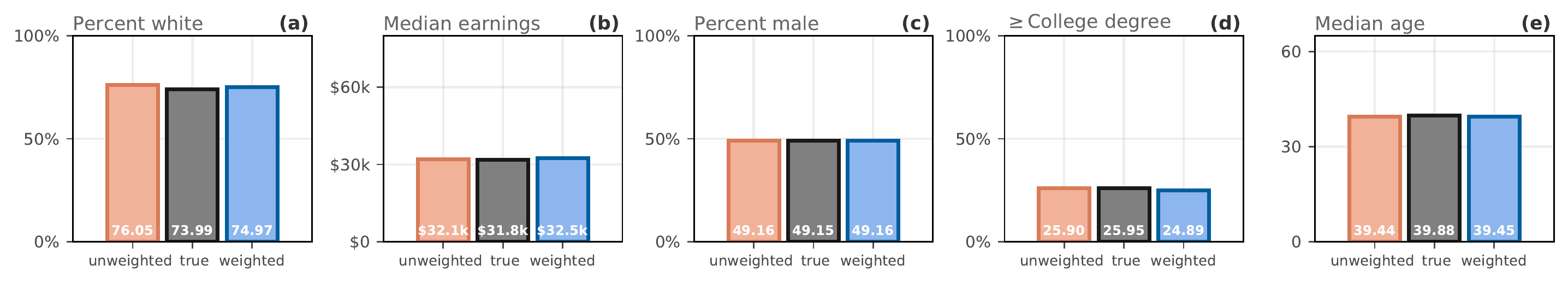}
    \end{subfigure}
    \caption{\textbf{National Level Socio-Demographics.} True values for five national level socio-demographic characteristics -- the proportion of males, median earnings, the proportion of having a college degree or higher, and a proportion of white users -- as recorded by the 2014-2018 5-year American Community Survey (ACS) data and their aggregated reconstructed values when considering the unweighted and weighted versions of the Panel of users.}
    \label{fig:demo_national}
\end{figure}

\begin{figure}[ht!]
    \centering
    \begin{subfigure}{0.95\columnwidth}
        \centering
        \includegraphics[width=1.0\columnwidth]{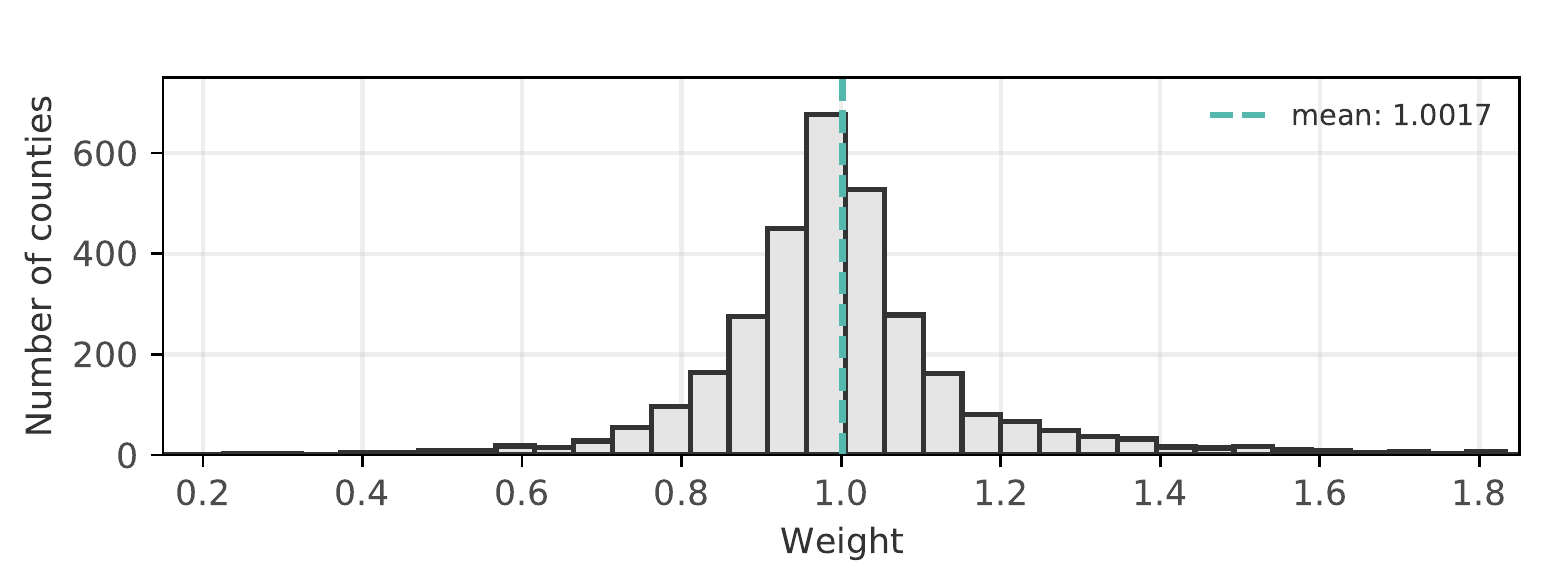}
    \end{subfigure}
    \caption{\textbf{County-level weights.} Distribution of county-specific sampling weights $\bar{w}_c$'s.}
    \label{fig:final_weights}
\end{figure}

Geolocation location data used in this study come from a panel of anonymized, opted-in devices from users in the United States. As modeling critical public health issues using a convenience sample of mobile users requires caution \cite{Buckee2020}, we checked the representativeness of our panel. In Figure \ref{fig:county_coverage}, we show that while the median coverage for our sample is only about $1.76\%$, the overall correlation between the population size of each county and the number of panel users per county is $96.7\%$. In addition, in Figure \ref{fig:demo_national}, we show that both the unweighted and weighted versions of our panel display a high degree of agreement with the values of the statistics as recorded by he 2014-2018 5-year American Community Survey (ACS). The weighted version of the panel has been developed with the aim of reducing any potential user selection bias in the data using an approach grounded in well-established statistical techniques \cite{curtis2007using,haneuse2009adjustment}.  

\begin{figure}[t!]
    \centering
    \begin{subfigure}{0.95\columnwidth}
        \centering
        \includegraphics[width=1.0\columnwidth]{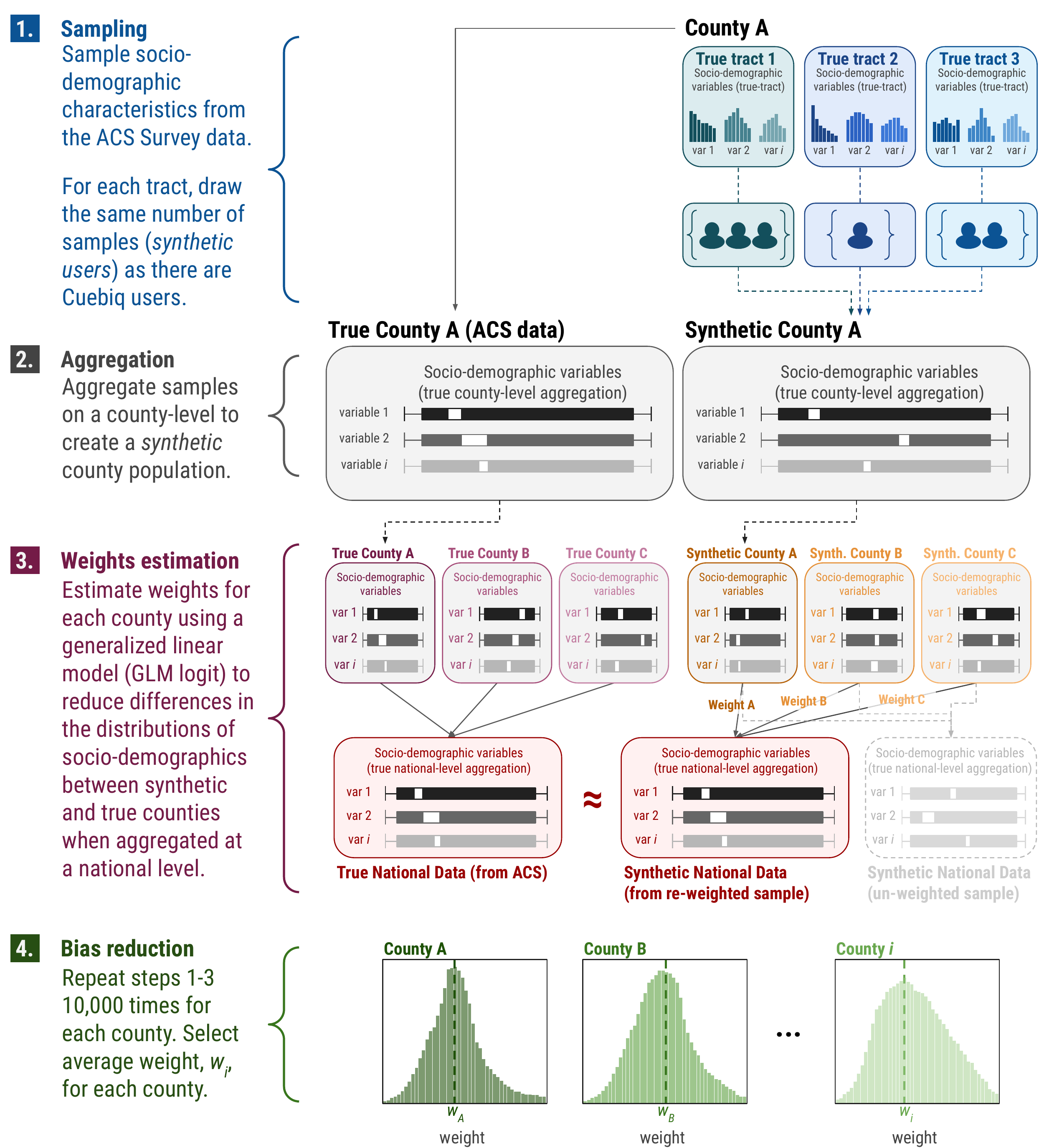}
    \end{subfigure}
    \caption{\textbf{Schematic of statistical procedure for assigning county-level weights.} Through repeatedly simulating synthetic populations at the \textit{census tract} level (based on the number of Cuebiq users with ``home'' personal areas in each census tract), we assign weights to the \textit{county} level in such a way that minimizes the bias with respect to demographic variables of interest. After 10,000 simulations, we select the average weight for each county.}
    \label{fig:reweighting_tmp}
\end{figure}

In particular, to reduce the selection bias that might exist at a sub-national level in terms of age, sex, race, educational attainment, and earnings (``demographics'' or ``socio-demographic characteristics''), we adjusted all measures presented in this manuscript using sampling weights, which allow us to create more statistically representative and, thus, more generalizable inferences from our data at the state and national levels of aggregation. Specifically, when aggregating the indexes at the state or national resolutions, we weigh the mobility and proximity metrics computed for each county, e.g., commutes, not only by the number of users present in our panel for a given location but also by a weight that corrects for the potential selection bias resulting from under- or over-sampling of users with certain socio-demographic characteristics. The weights are estimated using the method outlined in \cite{haneuse2009adjustment}, which allows us to estimate the parameters of a target population using data from a potentially biased sample, provided that the determinants of the selection bias are available for both the target population and the sample. In the context of this study, we are estimating mobility and contact patterns of the US population (i.e., the target population) using our panel data. We are assuming that users’ socio-demographics characteristics influence their probability of inclusion in the Cuebiq sample. Therefore, we use information about the distribution of these demographics in both the U.S.~population and our panel of users to compute bias-reducing weights. The schematic of the adopted statistical procedure is provided in Figure \ref{fig:reweighting_tmp} and described in the following.

First, we associate each user to a Census tract by using the location of their home personal area (see Section \ref{personal_areas_classification} in Materials \& Methods). This allows us to assign to each user a probability distribution of their socio-demographic characteristics by looking at their empirical distributions as reported for each Census tract in the 2014-2018 5-year American Community Survey (ACS) data \cite{ACS_data}. Second, we create a synthetic population of users in each census tract, which is a sub-unit of a county, using tract-level data from the ACS. The synthetic population’s size in each tract is determined by the number of panel users that are assigned to each tract. Then, to each user we randomly assign age, sex, race, educational attainment, and earnings by sampling their values from the census data. After generating one synthetic population, we compute the mean age, the proportion of males, mean earnings, the proportion of having a college degree or higher, and a proportion of white users for each synthetic county. This process is then repeated 10,000 times, therefore generating 10,000 synthetic datasets.  Third, we use a generalized linear model (GLM) with a binomial link function (logit) to estimate the probability that a given county is a synthetic county (as generated from the sample of panel users) or a ``census'' county where for the given county we directly use the census mean values for the different indicators. In other words, we treat this problem as a classification problem where our regression uses the computed county-specific socio-demographic summary statistics to predict whether a county has been simulated using the Cuebiq sample or not. The intuition behind this procedure is that if our sample is unbiased in terms of the demographics, then the demographic information should not allow us to predict whether a county is a synthetic county or a ``census'' county (in which case, all estimated $\beta$ coefficients should not be statistically significantly different from zero). Conversely, if the sample is biased, demographics will produce meaningful predictions.

Using this family of GLMs in the process of reducing sampling bias is standard in well-established techniques such as inverse probability weighting and propensity score matching  \cite{curtis2007using,dehejia2002propensity}. To estimate county-specific weights that reduce bias at the national level, accounting for state effects, we use the following model specification:
\begin{equation} \label{eq:3}
\begin{split}
    P({\text{synthetic}}) =& \beta_1 \times \text{age} + \beta_2 \times \text{college} + \beta_3 \times \text{college}^2 + \\ & \beta_4 \times \text{earnings} + \beta_5 \times \text{white} + \beta_6 \times \text{male} + \{\text{state}_s\}_{s=1}^{50};
\end{split}
\end{equation}
where $age$ denotes the county mean age, $college$ denotes the proportion of the total population having a college degree or higher, $white$ denotes the proportion of the total population being white, $earnings$ denotes the average earnings, $male$ denotes the proportion of males in a county, and $state_s$ is a state-specific fixed effect. All variables are z-score standardized. 

Lastly, we fit our statistical models to each one of the 10,000 synthetic datasets, compute the probability that a given county is a ``synthetic'' county ($p_{\text{synthetic}}$), and convert it into a county-specific weight $w^i_{c}$ using:
\begin{equation}
    w^i_{c} = \frac{1}{p_{\text{synthetic}}} - 1,
\end{equation}
where $i$ denotes the dataset used and $c$ denotes the county \cite{haneuse2009adjustment}. We then obtain our final county-specific weight $\bar{w}_c$ as the average of all the estimated weights: $\bar{w}_c = \sum_{i=1}^{I} w^i_c/I$ where $I=$10,000 (see Figure \ref{fig:final_weights}).

\subsubsection{Sensitivity Analysis}\label{sec:si_sensitivity}

\begin{figure}[t!]
    \centering
    \begin{subfigure}{0.95\columnwidth}
        \centering
        \includegraphics[width=1.0\columnwidth]{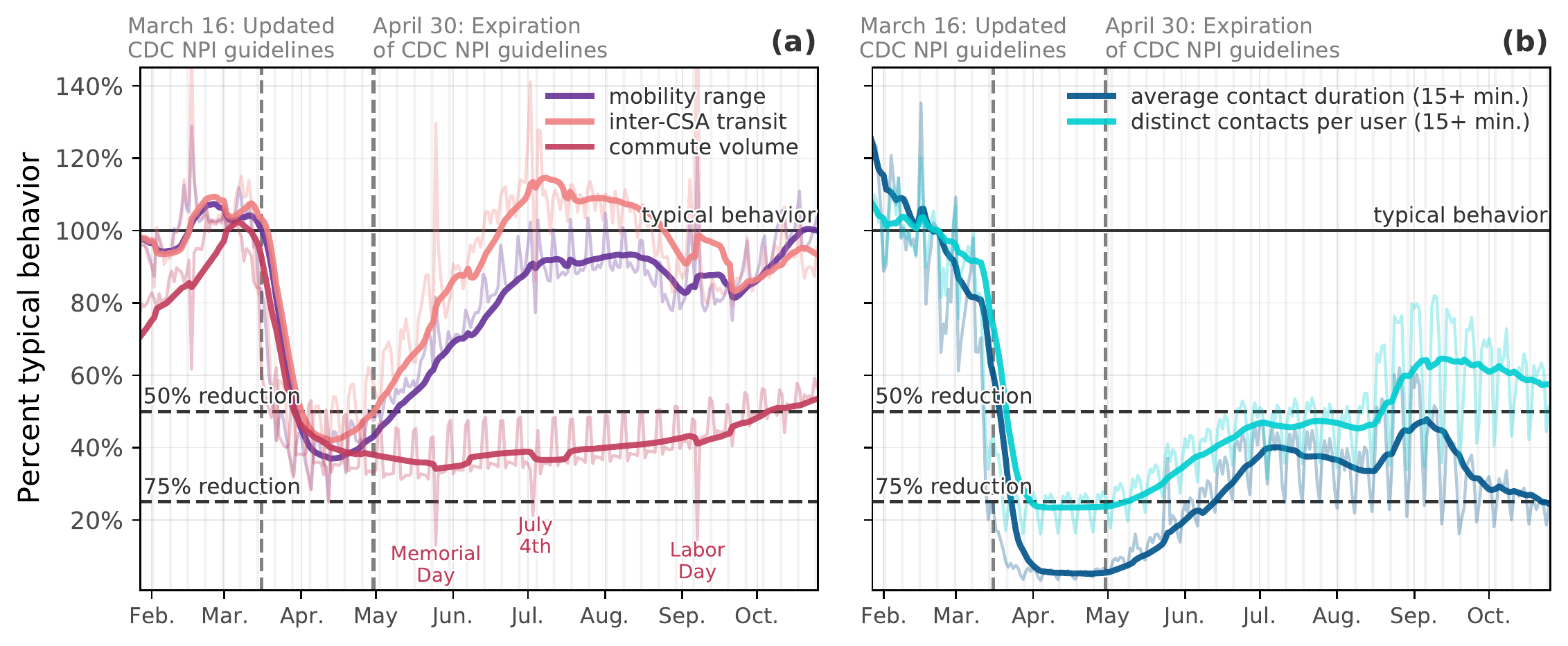}
    \end{subfigure}
    \caption{\textbf{Changes in mobility and person-to-person contacts over time (unweighted panel).} Graphs show deviations from typical behavior for the same weekday, in the United States, using the \textit{unweighted} panel. \textbf{(a)} Mobility: Individual mobility (radius of gyration), commute volume, and inter-CSA transit. \textbf{(b)} Contacts: Number of distinct contacts and average contact duration events outside of work and home. By the national declaration of emergency (March 13), reductions in spatial mobility measures had begun, reaching approximately $50\%$ of typical values by April 1; while contact measures show a reduction greater than $75\%$ by the same date. A 7-day rolling average is shown alongside each measure. Grey vertical lines denote weekends.}
        \label{fig:changes_unweighted}
\end{figure}

\begin{figure}[t!]
    \centering
    \begin{subfigure}{1.0\columnwidth}
        \centering
        \includegraphics[width=1.0\columnwidth]{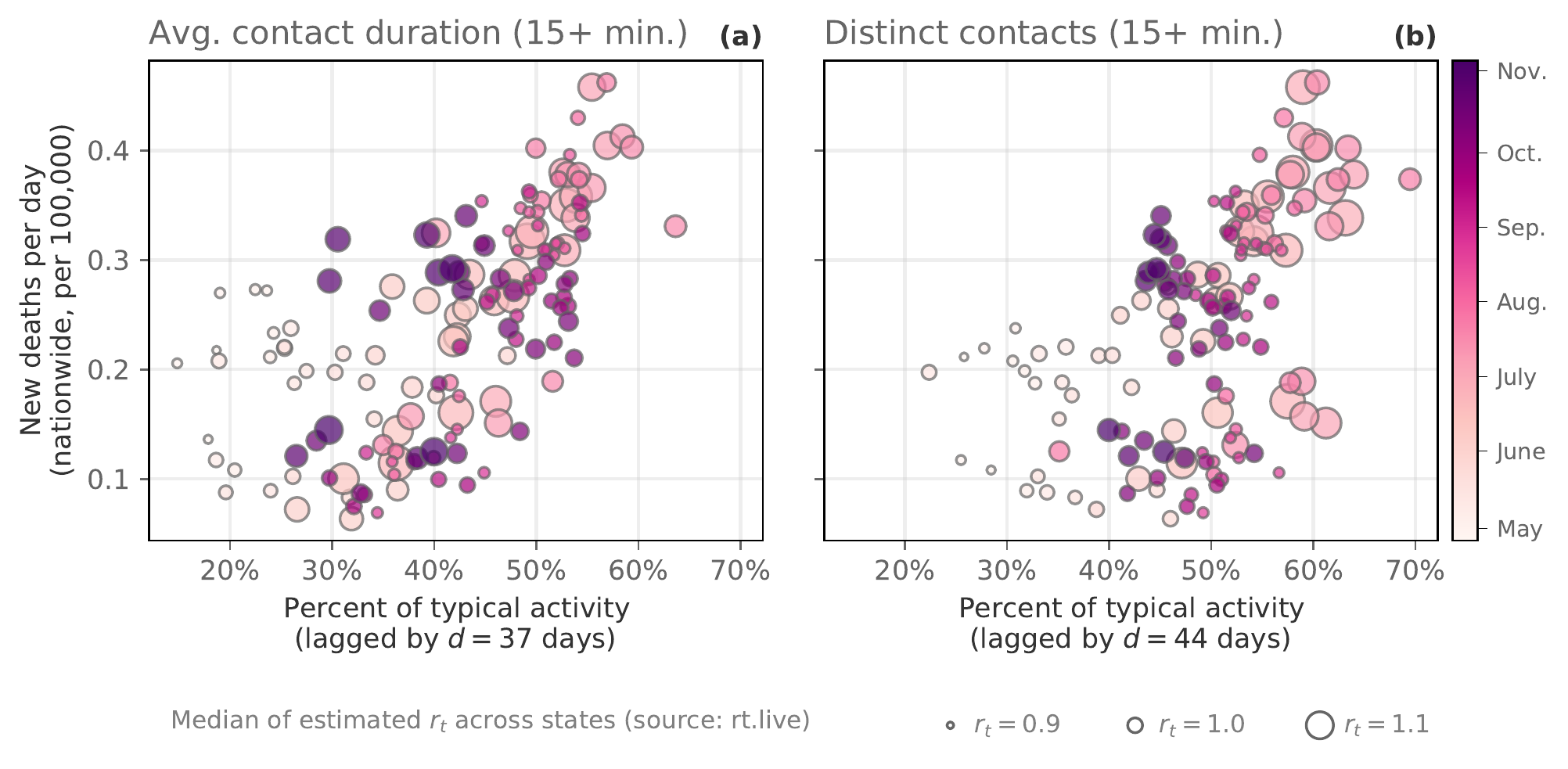}
    \end{subfigure}
    \caption{\textbf{Collective physical distancing and new deaths (unweighted panel).} Replication of the analysis in Figure \ref{fig:newdeaths} using the unweighted panel to correlate daily contact measures nationwide with new reported deaths \cite{CovidTracking} between April 30 and November 5, 2020.}
    \label{fig:newdeaths_unweighted}
\end{figure}

To evaluate the effects of the weighted resampling procedure, here we replicate some of the results present in the main text using the unweighted panel. In particular, in Figure \ref{fig:changes_unweighted} we show the changes in mobility and contacts over time, and in Figure \ref{fig:newdeaths_unweighted} we report the correlation between our measures of collective physical distancing and new deaths. In both cases, results are in line with the ones obtained using the weighted sampling procedure.

\begin{figure}[ht!]
    \centering
    \includegraphics[width=1.0\columnwidth]{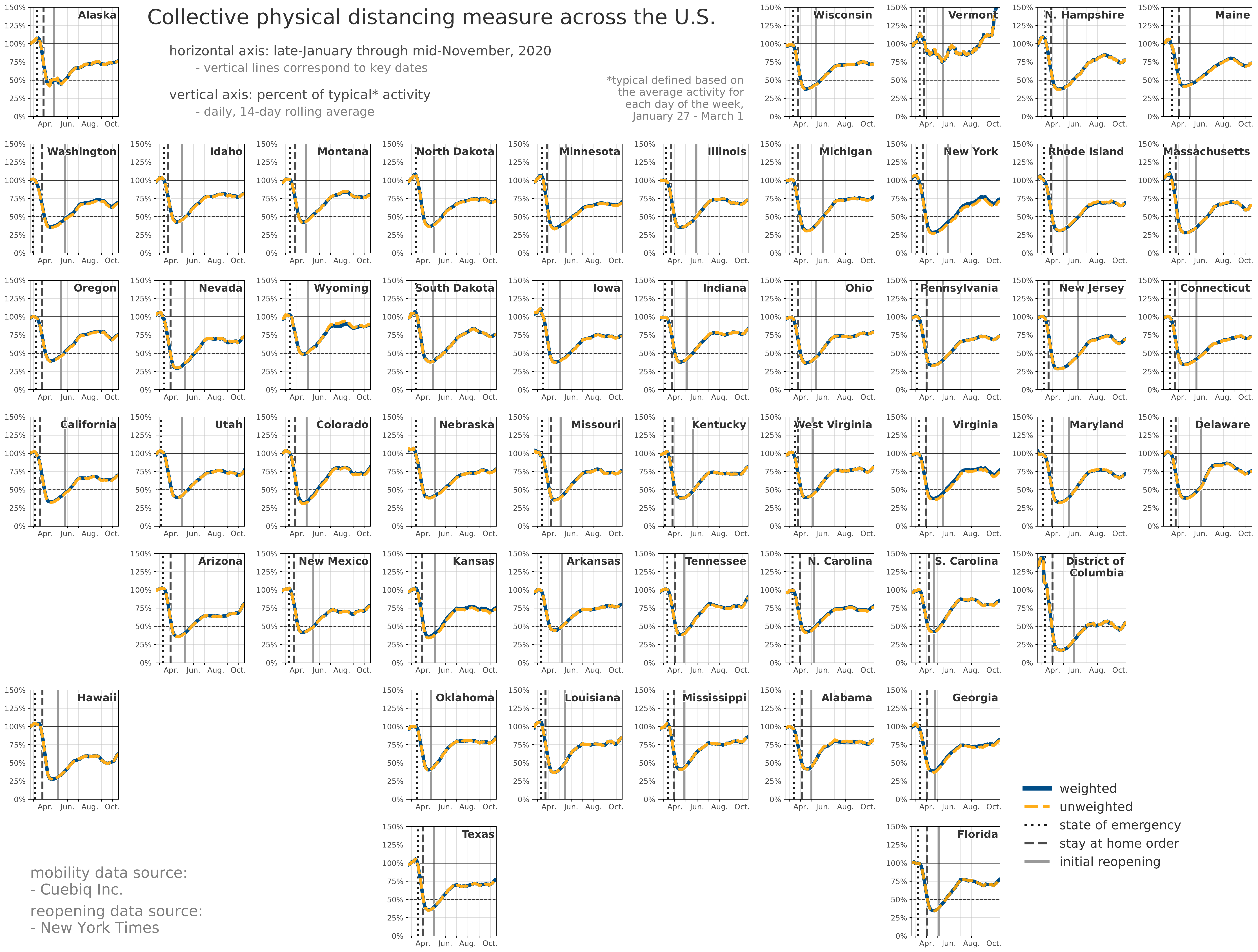}
    \caption{\textbf{Collective Physical Distancing: Weighted vs unweighted.} For each state we report the aggregate measure of collective physical distancing, defined as the average of the typical daily commute volume, individual mobility range, inter-CSA transit, unique contacts outside of home and work, and total duration of contacts for the time frame listed, using the weighted panel (solid line) and the unweighted panel (dashed line).}
    \label{fig:map_collective}
\end{figure}

In Figure \ref{fig:map_collective}, instead, we show the effect of using the unweighted vs the weighted panel in computing the time series of the average collective physical distancing (as defined in Figure \ref{fig:aggregated_reduction_states_maps}) for each state. While there are slight differences in some states, the overall picture and insights from our analysis do not change.

\begin{figure}[t!]
    \centering
    \begin{subfigure}{0.95\columnwidth}
        \centering
        \includegraphics[width=1.0\columnwidth]{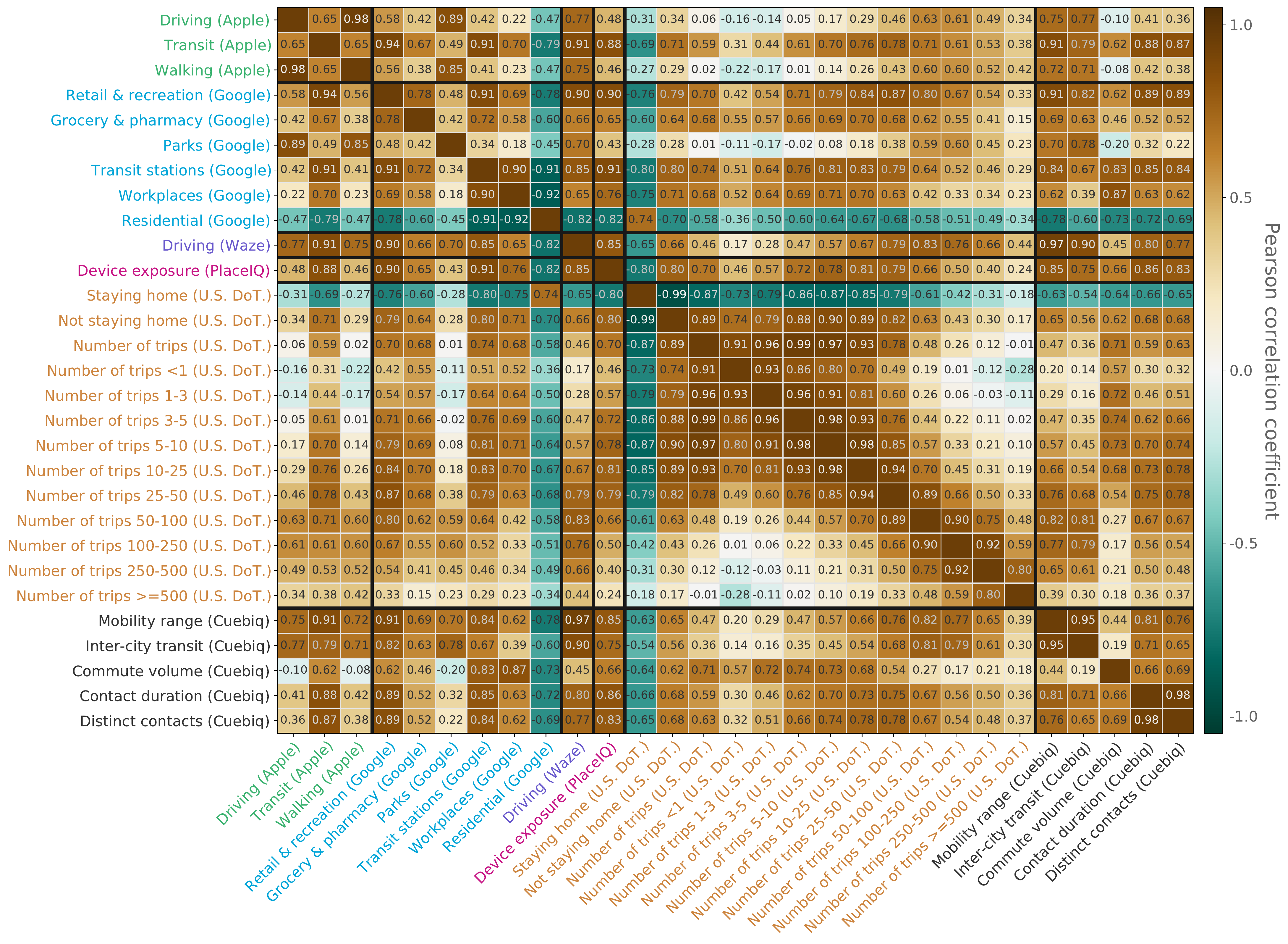}
    \end{subfigure}
    \caption{\textbf{Correlations across mobility datasets.} Correlation matrix comparing the daily time series of mobility and contact patterns across different datasets.}
    \label{fig:comparison_heatmap2}
\end{figure}

\begin{figure}[t!]
    \centering
    \includegraphics[width=0.9\columnwidth]{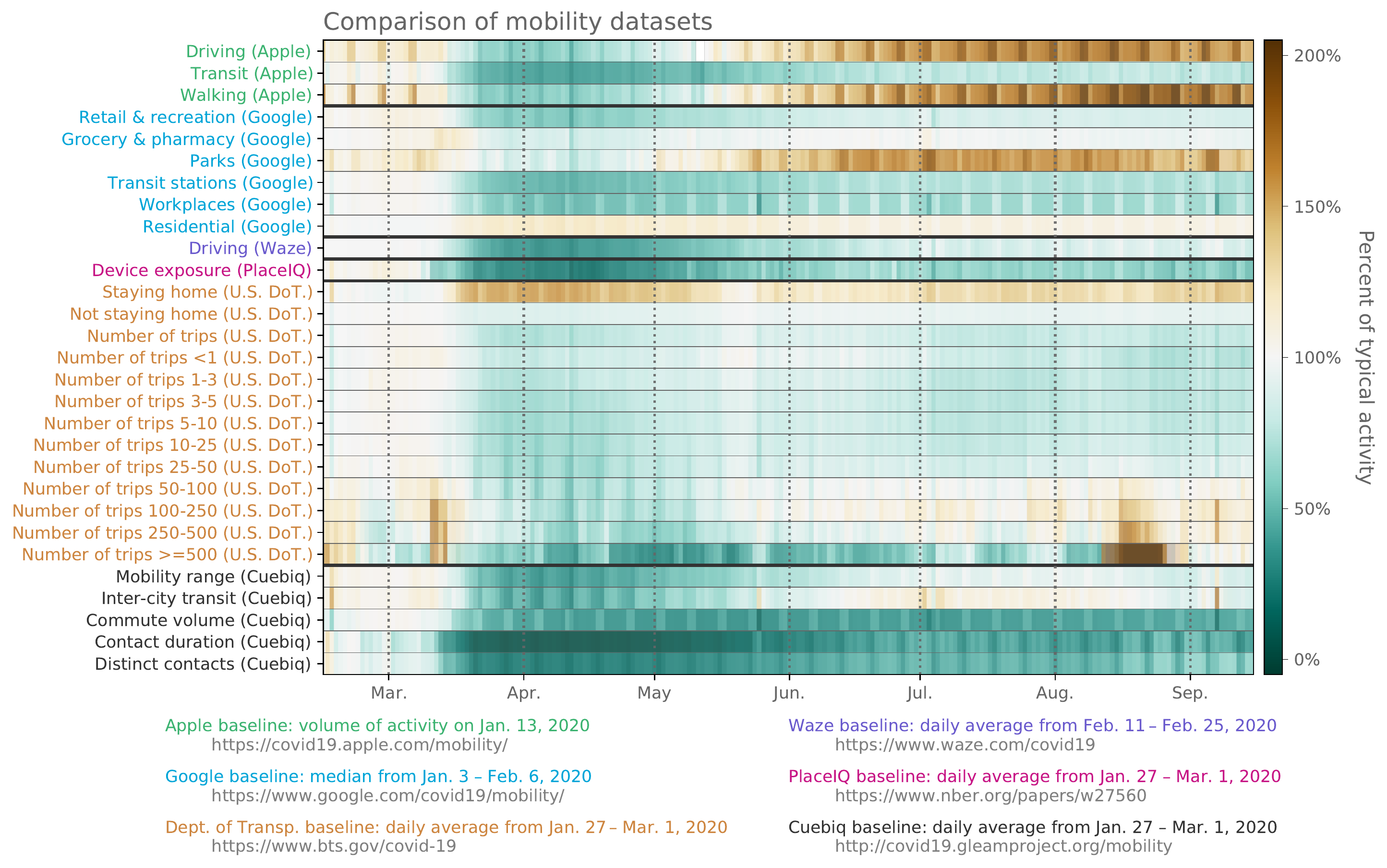}
    \caption{\textbf{Comparison across mobility datasets.} While not meant to be a comprehensive comparison, this heatmap offers a coarse estimate of the similarities and differences of a variety of mobility measures from different datasets.}
    \label{fig:comparison_heatmap}
\end{figure}

\subsection{Correlating physical distancing measures across datasets}

In this work, we use data from Cuebiq Inc., but one feature of the COVID-19 pandemic is that mobile providers and other large technology companies have been providing access to aggregated measures of mobility and contacts. For this reason, we include here a series of correlations between the measures studied here and those from a number of other platforms. As a proof-of-principle validation, the measures we include strike a key balance between correlating with existing publicly-available mobility measures (e.g. Google's ``residential'' measure negatively correlates with each of our measures---which makes sense, as we do not use location pings from within users' home locations) and still providing unique information. The datasets included in Figures \ref{fig:comparison_heatmap} and \ref{fig:comparison_heatmap2} are from: Google (\url{https://www.google.com/covid19/mobility/}), Apple (\url{https://covid19.apple.com/mobility/}), PlaceIQ (\url{https://www.nber.org/papers/w27560}), Waze (\url{https://www.waze.com/covid19}), and the U.S.~Bureau of Transportation Statistics (\url{https://www.bts.gov/covid-19}).

Broadly, there is correspondence between the measures introduced here and those used by Apple, Google, PlaceIQ, and the U.S.~Dept. of Transportation (Figure \ref{fig:comparison_heatmap2}). The measures that we expect to be highly correlated are indeed highly correlated: for example, Google's ``workplace'' measure and our commute volume are Pearson correlated at 0.87. Similarly, our mobility measure is highly correlated with Google's and Apple's ``transit'' measures, and it is negatively correlated with Google's ``residential'' measure.

Another point of validation can be seen when comparing the various time series of activity (Figure \ref{fig:comparison_heatmap}). For major holidays, where we would expect movement to be disrupted, we see broad alignment between the various measures. For example, in early September (Labor Day), we see an increase in the Dept.~of Transportation's ``Number of trips 100 miles+'' measures; similarly we see the same spike in our inter-city transit measure.

There are endless ways to compare the myriad measures of mobility that have been studied during the COVID-19 pandemic, and despite this, the measures included in this work are balanced between offering a novel, informative lens to understand collective physical distancing while also corresponding neatly to measures that have already been proposed in the literature.

\subsection{Correlating contact patterns with new positive tests}

\begin{figure}[t!]
    \centering
    \includegraphics[width=0.775\columnwidth]{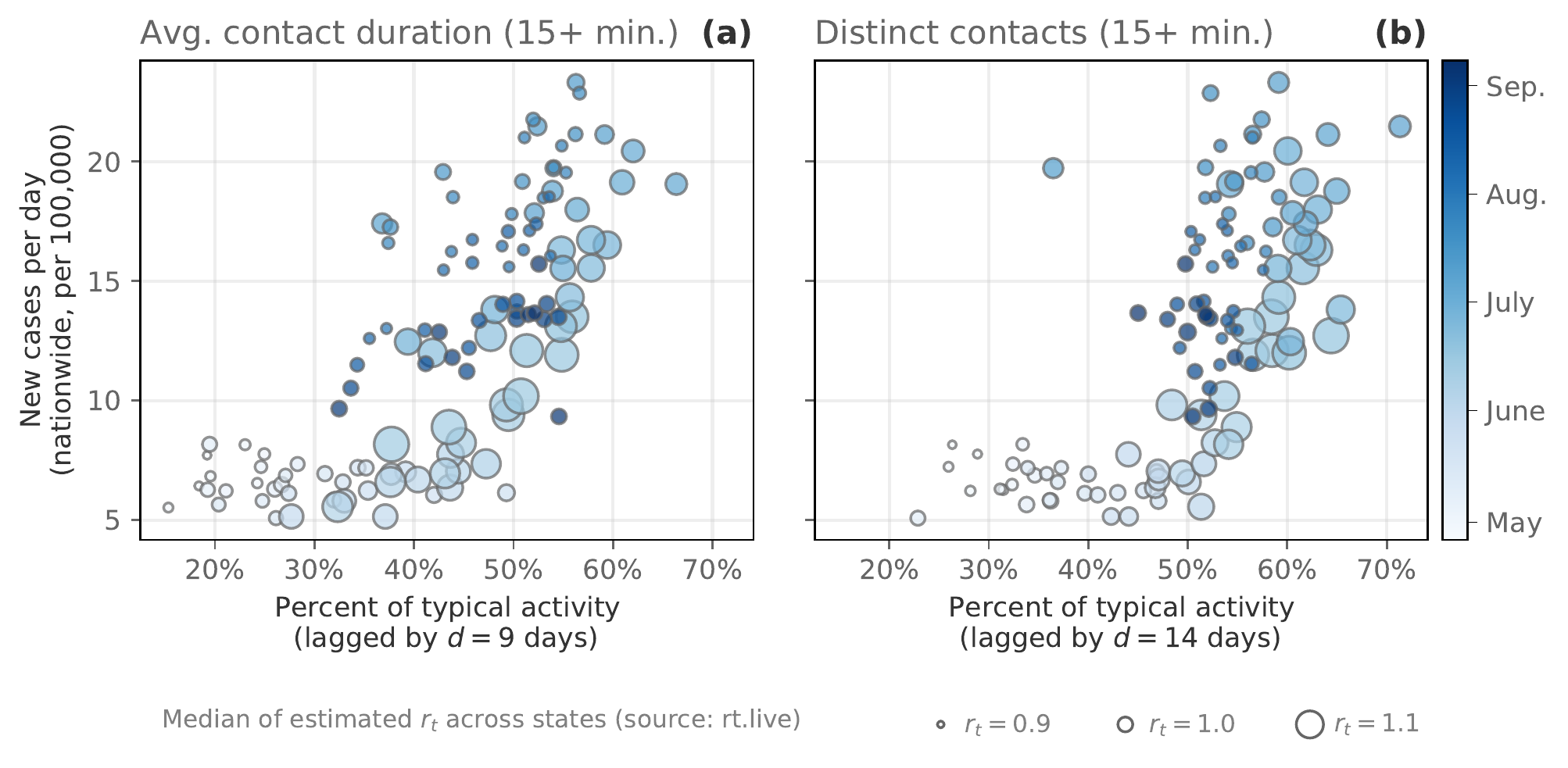}
    \caption{\textbf{Collective physical distancing and new infections.} Correlating contact measures nationwide (lagged) with new reported cases daily per 100,000 (data from \textit{The COVID Tracking Project} \cite{CovidTracking}) between April 30 and September 5, 2020. A lag of $d$ days was selected for each state so as to maximize the $R$-squared of the correlation between average contact duration and new infections. \textbf{(a)} Average contact duration \textbf{(b)} Distinct contacts. In each subplot, darker colors indicate later dates and marker size corresponds to an estimate of the median $R_t$ across all 50 states and District of Columbia (source: \url{rt.live}).}
    \label{fig:newcases}
\end{figure}

The contact measures introduced in this manuscript are meant to be used as a lower resolution approximation for true person-to-person contacts in the U.S.\ population. As we showed in Section \ref{sec:new_deaths}, the average contact duration and distinct contact measures both are positively correlated with (lagged) new reported deaths. In Figure \ref{fig:newcases}, we show that this pattern also holds for new positive tests at the national level and for most states (data from the COVID Tracking Project \cite{CovidTracking}). Again, we plot the lagged correlation that maximizes the $R$-squared; in this case, the delay that maximizes the average contact duration is 9 days, while distinct contacts is maximized at $d=14$ days. While testing data is typically noisier and depends on local testing policies, the presence of these positive correlations between contact patterns and lagged new cases is again suggestive that the measures used in this work can serve as coarse indicators of large-scale human behavior.

\clearpage
\subsection{Collective physical distancing at state and metropolitan levels} \label{sec:extra_figures}

\begin{figure}[ht]
    \centering
    \begin{subfigure}{0.95\columnwidth}
        \centering
        \includegraphics[width=1.0\columnwidth]{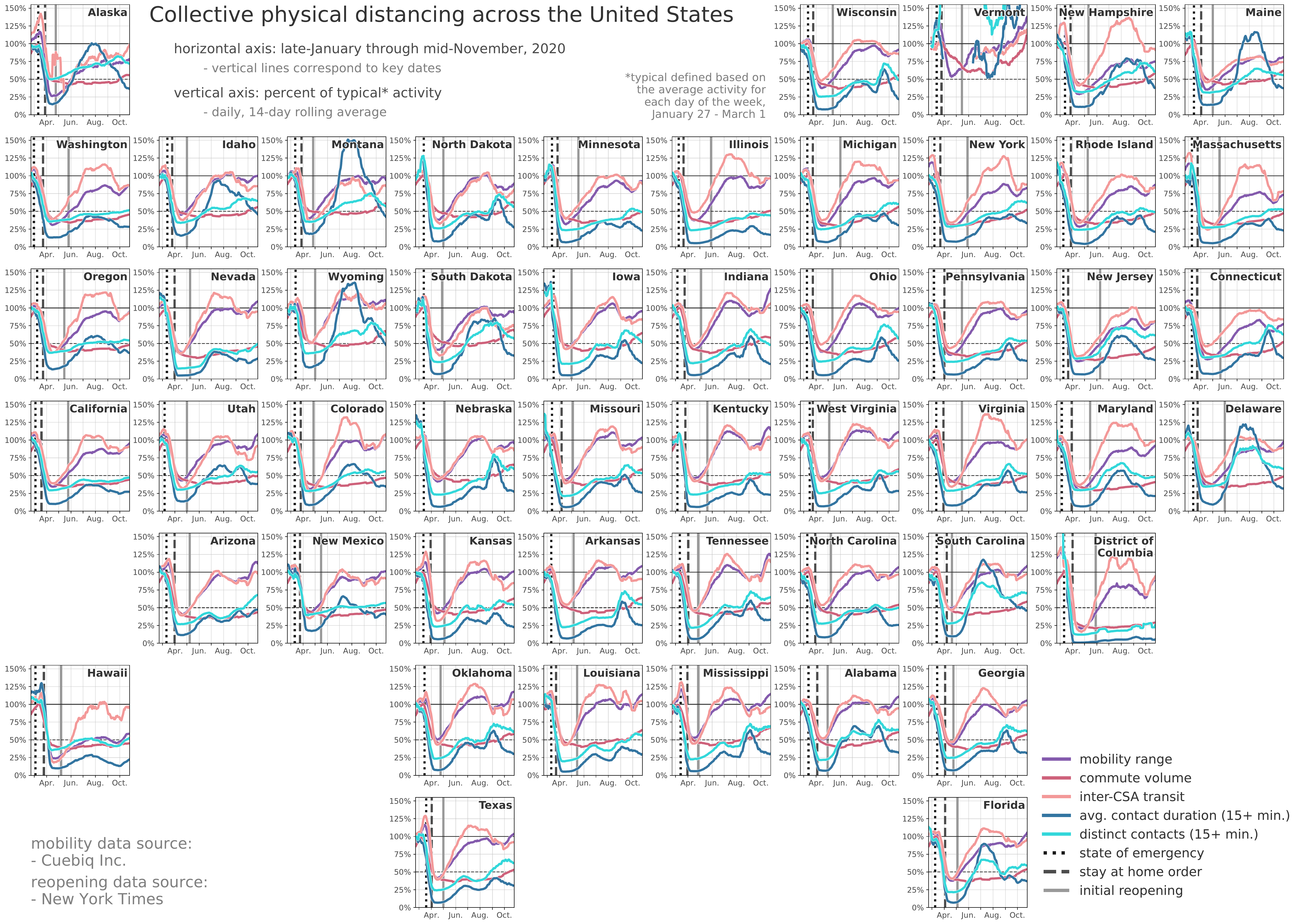}
    \end{subfigure}
    \caption{\textbf{Collective physical distancing across every state.} Grid cartogram including the five measures shown in Figure \ref{fig:five_measures_collapsed}, for all 50 states and District of Columbia.}
    \label{fig:allstates}
\end{figure}

\begin{figure}[ht]
    \centering
    \begin{subfigure}{0.95\columnwidth}
        \centering
        \includegraphics[width=1.0\columnwidth]{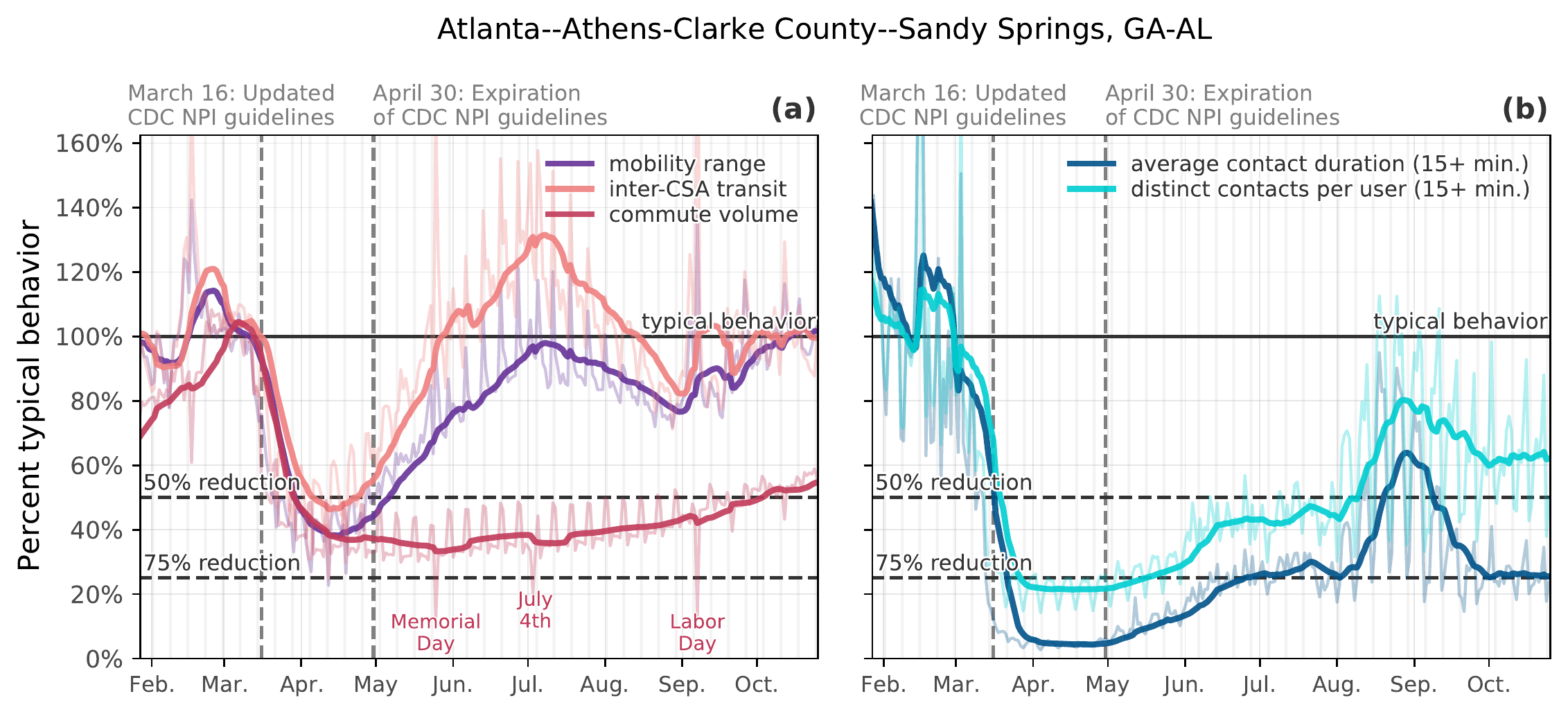}
    \end{subfigure}
    \caption{Atlanta--Athens-Clarke County--Sandy Springs, GA-AL.}
    \label{fig:city1}
\end{figure}

\begin{figure}[ht]
    \centering
    \begin{subfigure}{0.95\columnwidth}
        \centering
        \includegraphics[width=1.0\columnwidth]{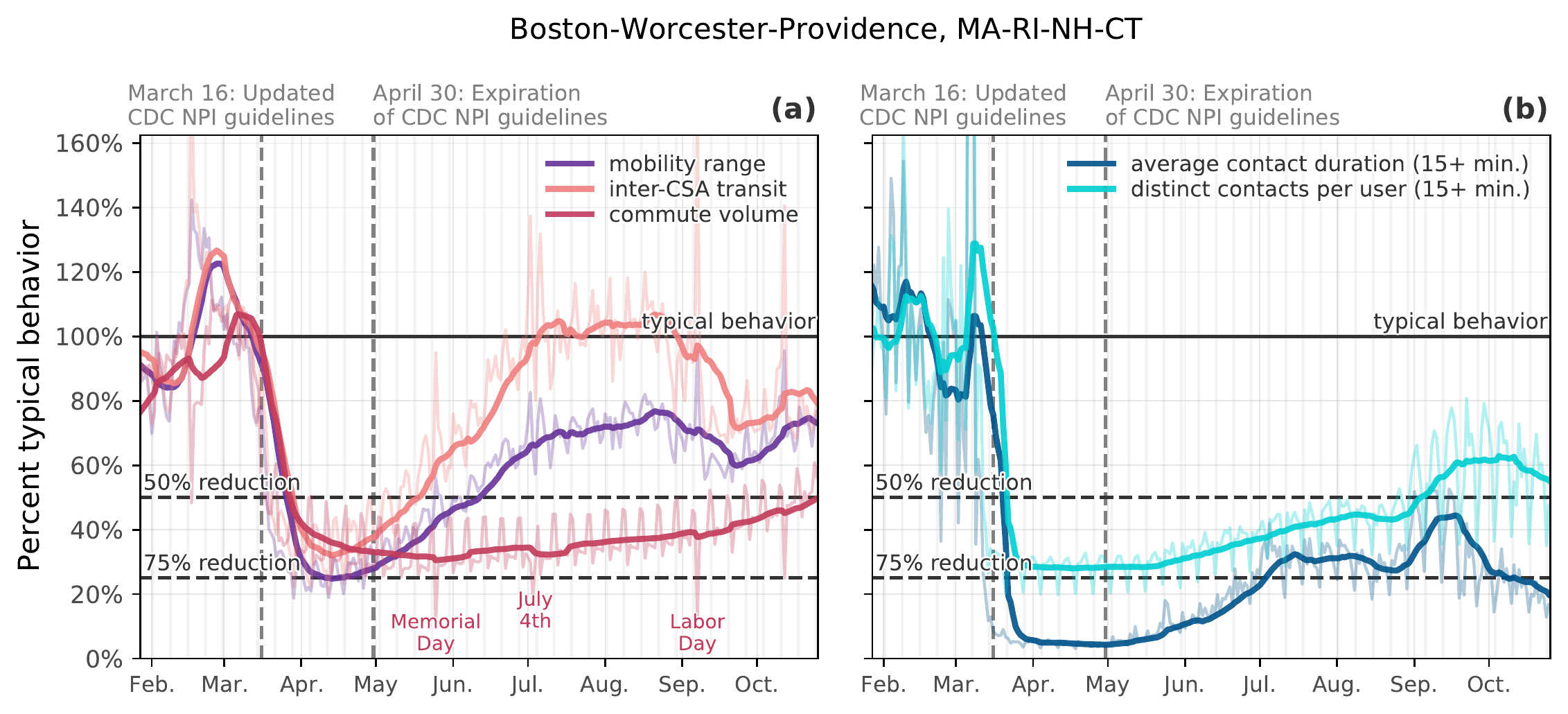}
    \end{subfigure}
    \caption{Boston-Worcester-Providence, MA-RI-NH-CT.}
    \label{fig:city2}
\end{figure}

\begin{figure}[ht]
    \centering
    \begin{subfigure}{0.95\columnwidth}
        \centering
        \includegraphics[width=1.0\columnwidth]{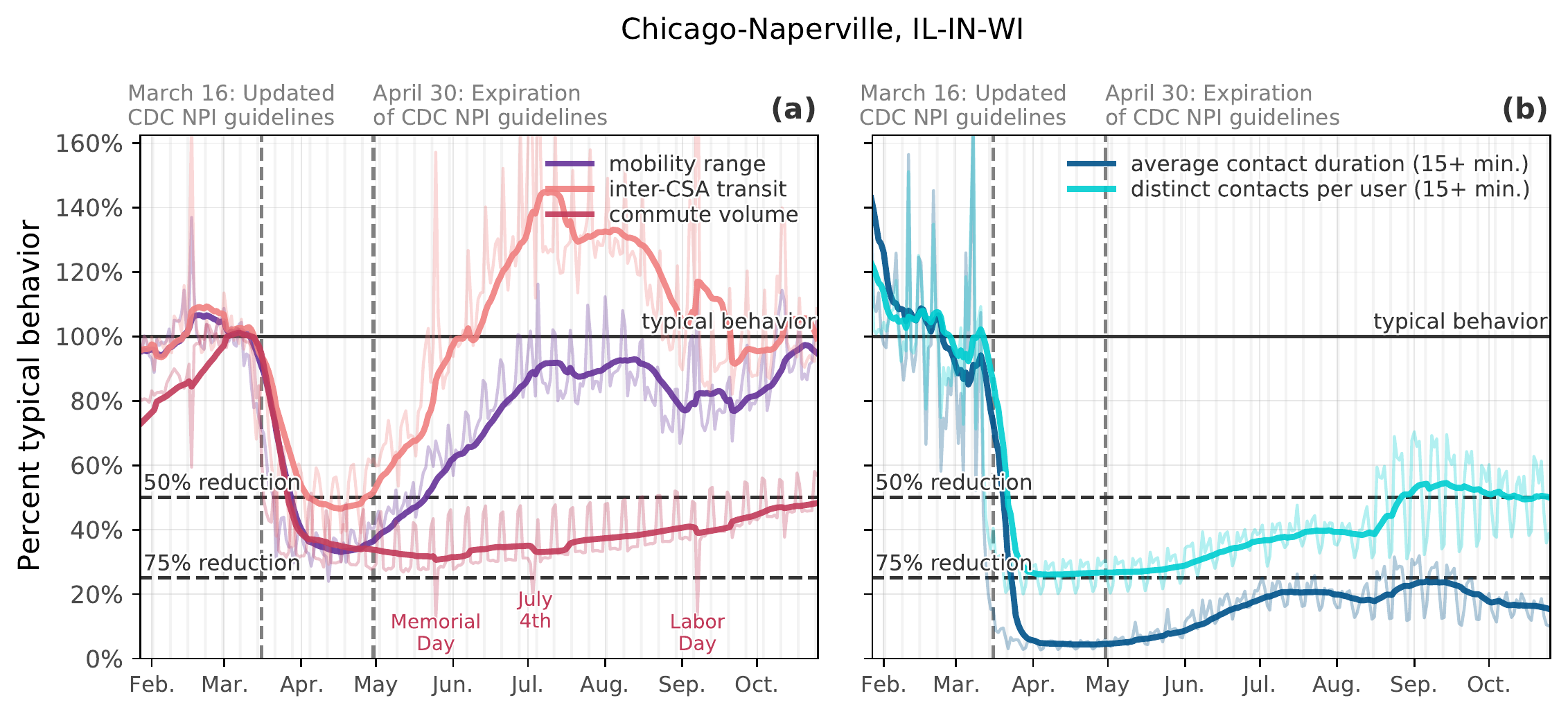}
    \end{subfigure}
    \caption{Chicago-Naperville, IL-IN-WI.}
    \label{fig:city3}
\end{figure}

\begin{figure}[ht]
    \centering
    \begin{subfigure}{0.95\columnwidth}
        \centering
        \includegraphics[width=1.0\columnwidth]{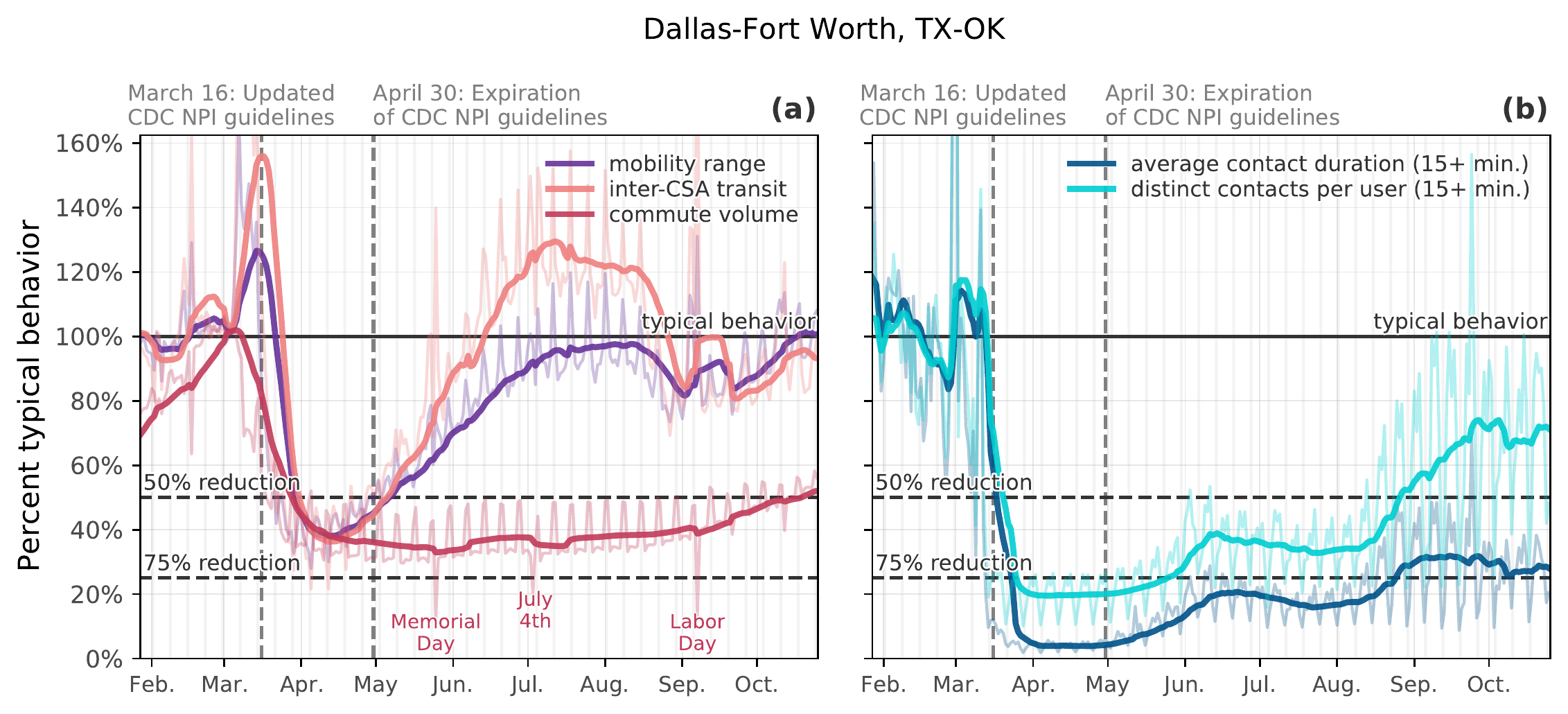}
    \end{subfigure}
    \caption{Dallas-Fort Worth, TX-OK.}
    \label{fig:city4}
\end{figure}

\begin{figure}[ht]
    \centering
    \begin{subfigure}{0.95\columnwidth}
        \centering
        \includegraphics[width=1.0\columnwidth]{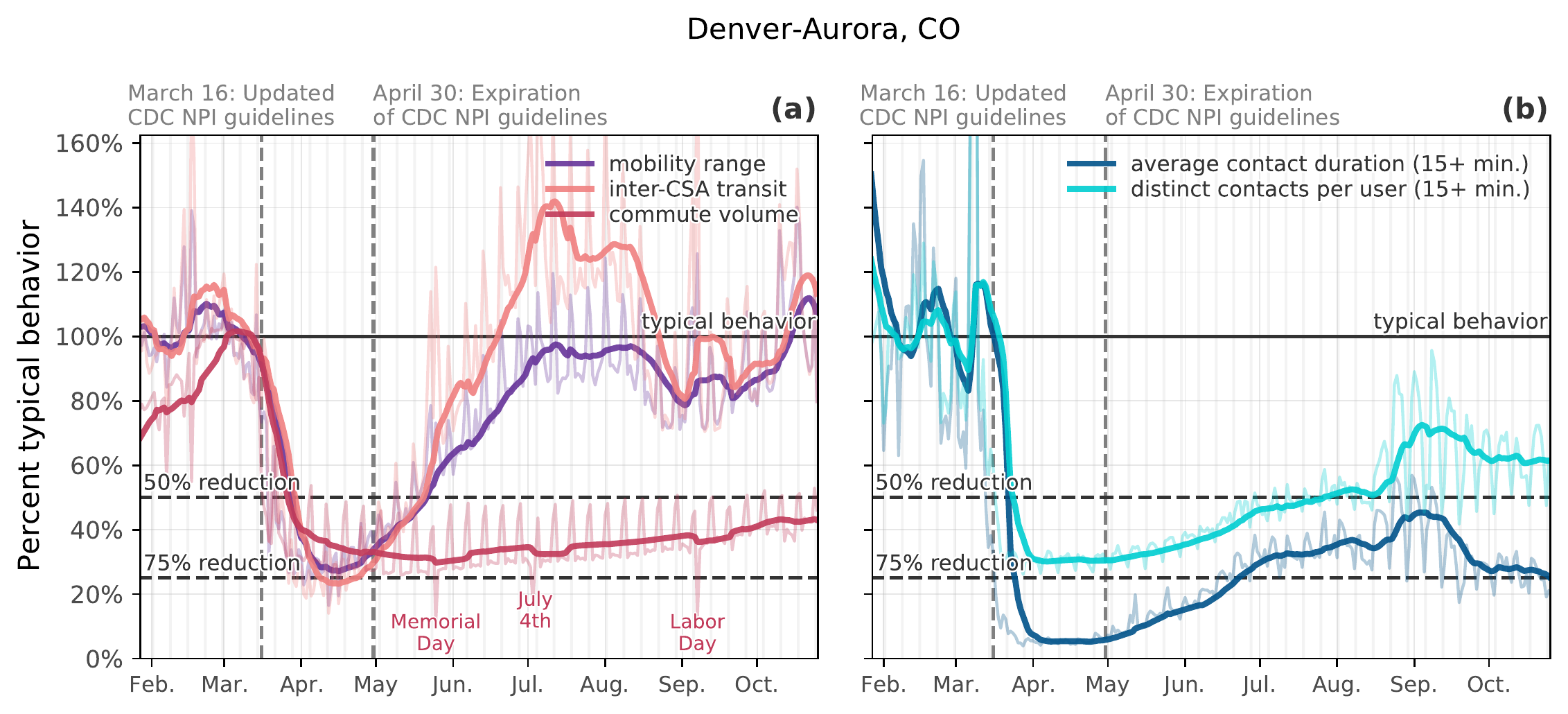}
    \end{subfigure}
    \caption{Denver-Aurora, CO.}
    \label{fig:city5}
\end{figure}

\begin{figure}[ht]
    \centering
    \begin{subfigure}{0.95\columnwidth}
        \centering
        \includegraphics[width=1.0\columnwidth]{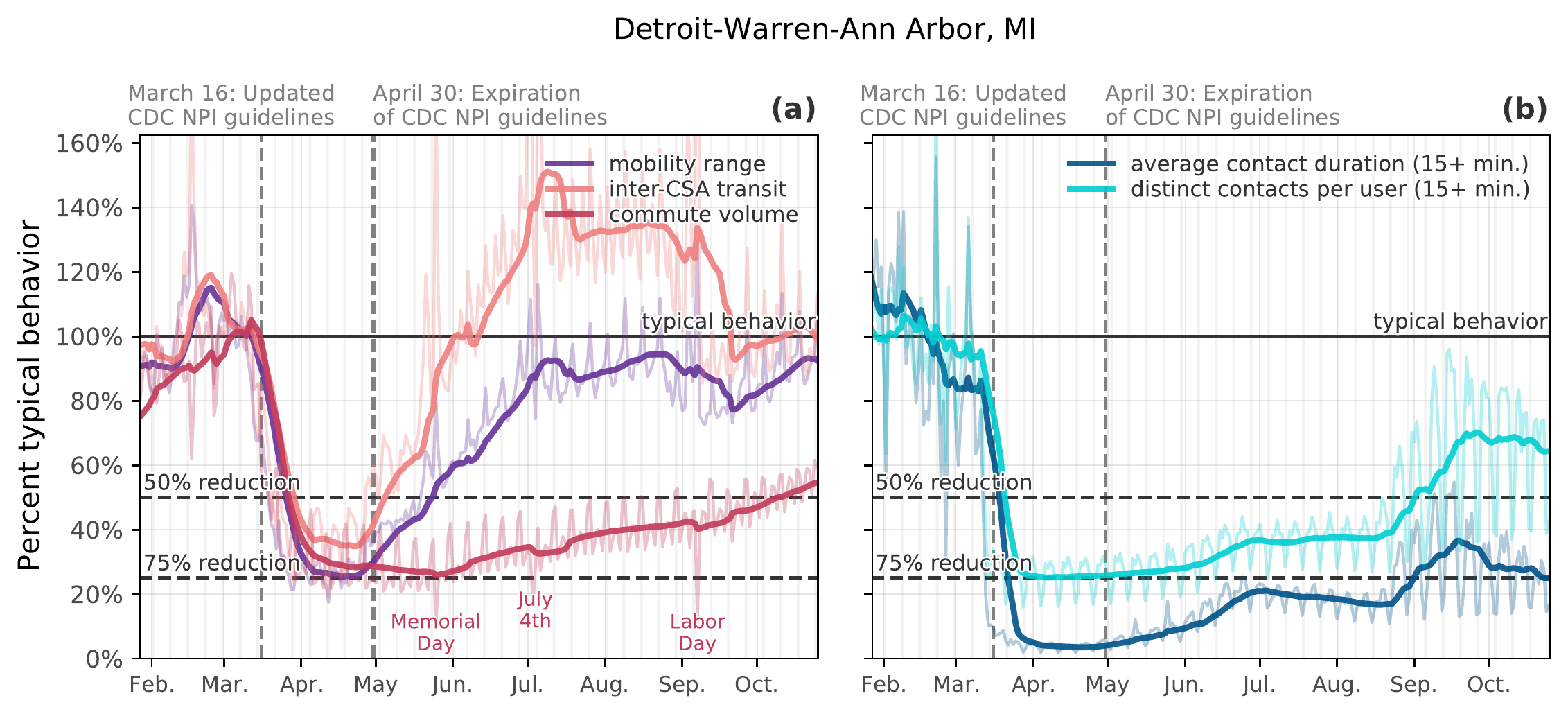}
    \end{subfigure}
    \caption{Detroit-Warren-Ann Arbor, MI.}
    \label{fig:city6}
\end{figure}

\begin{figure}[ht]
    \centering
    \begin{subfigure}{0.95\columnwidth}
        \centering
        \includegraphics[width=1.0\columnwidth]{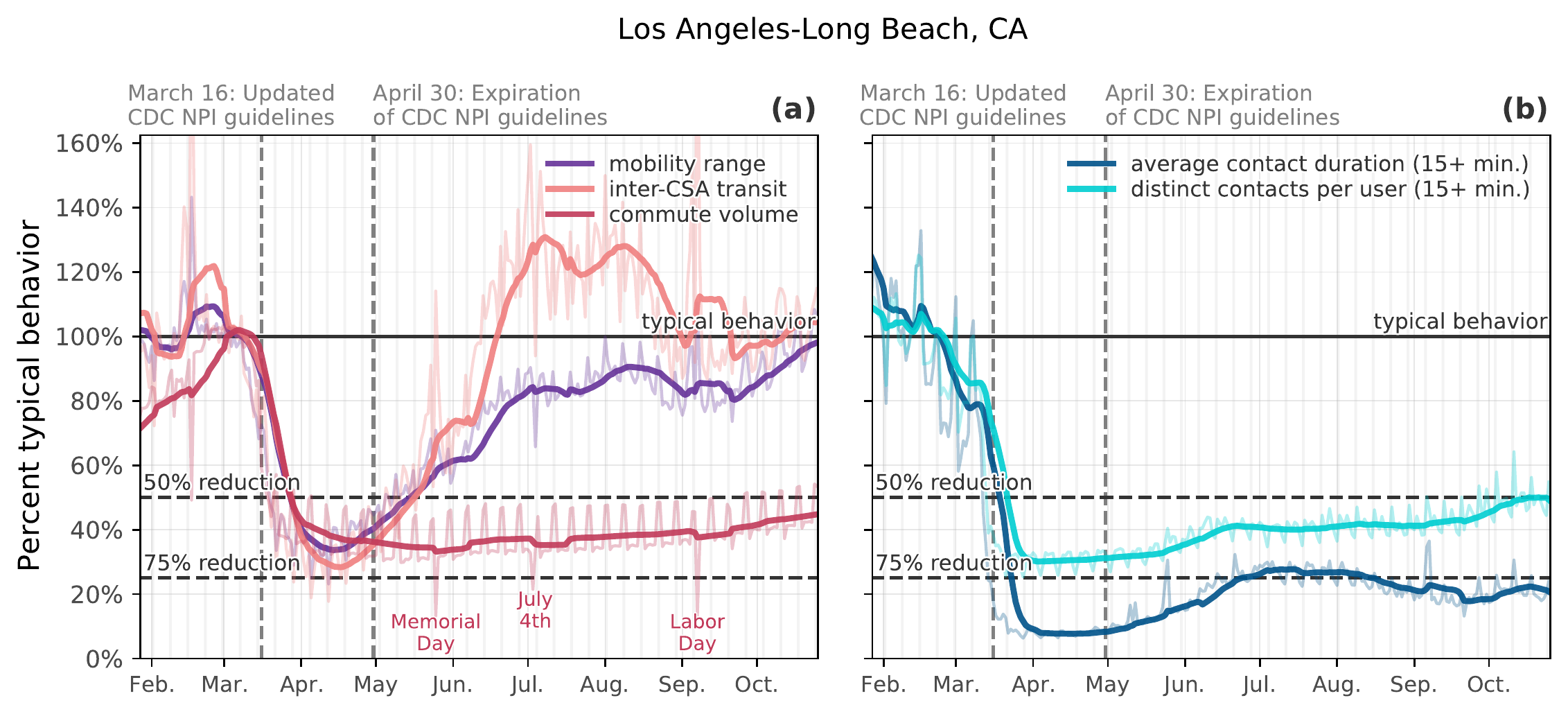}
    \end{subfigure}
    \caption{Los Angeles-Long Beach, CA.}
    \label{fig:city7}
\end{figure}

\begin{figure}[ht]
    \centering
    \begin{subfigure}{0.95\columnwidth}
        \centering
        \includegraphics[width=1.0\columnwidth]{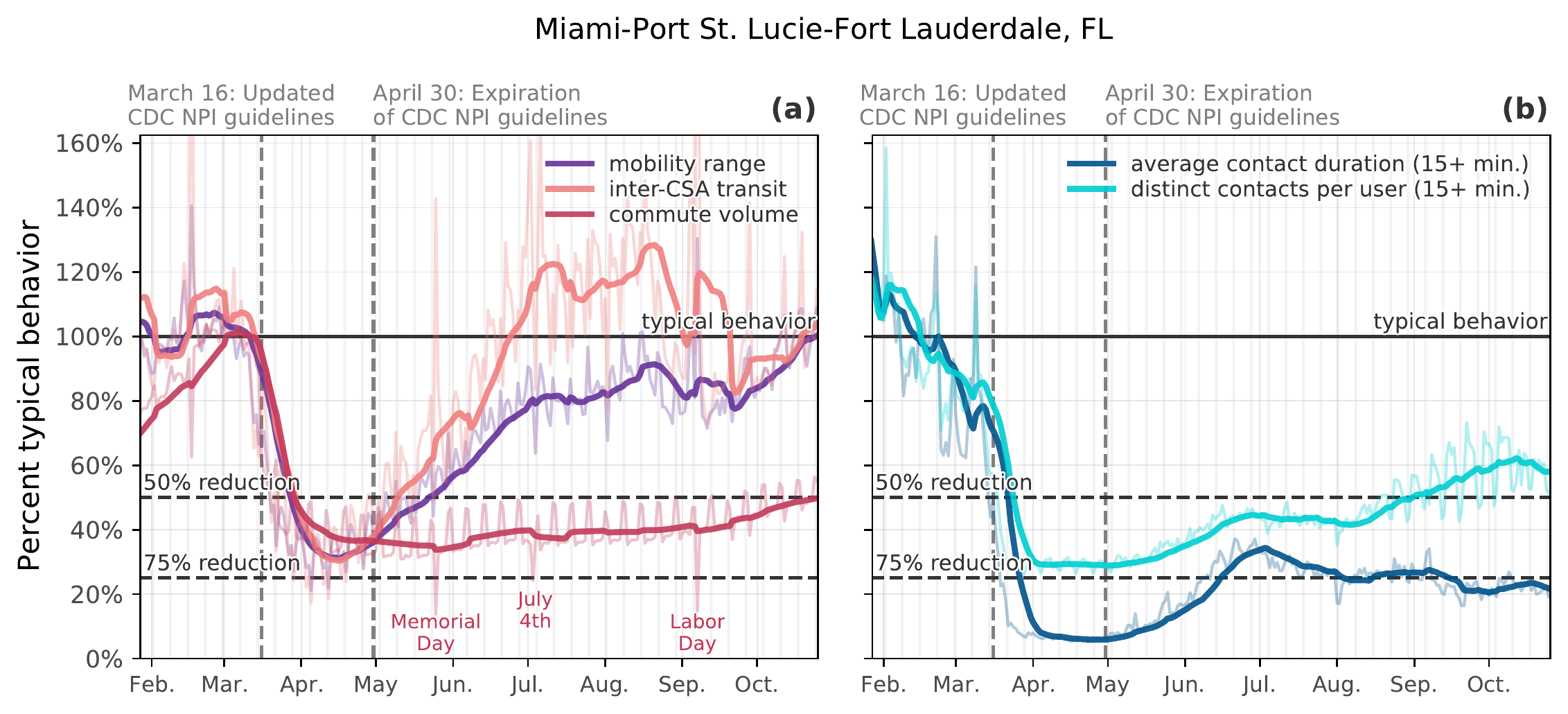}
    \end{subfigure}
    \caption{Miami-Port St. Lucie-Fort Lauderdale, FL.}
    \label{fig:city8}
\end{figure}

\begin{figure}[ht]
    \centering
    \begin{subfigure}{0.95\columnwidth}
        \centering
        \includegraphics[width=1.0\columnwidth]{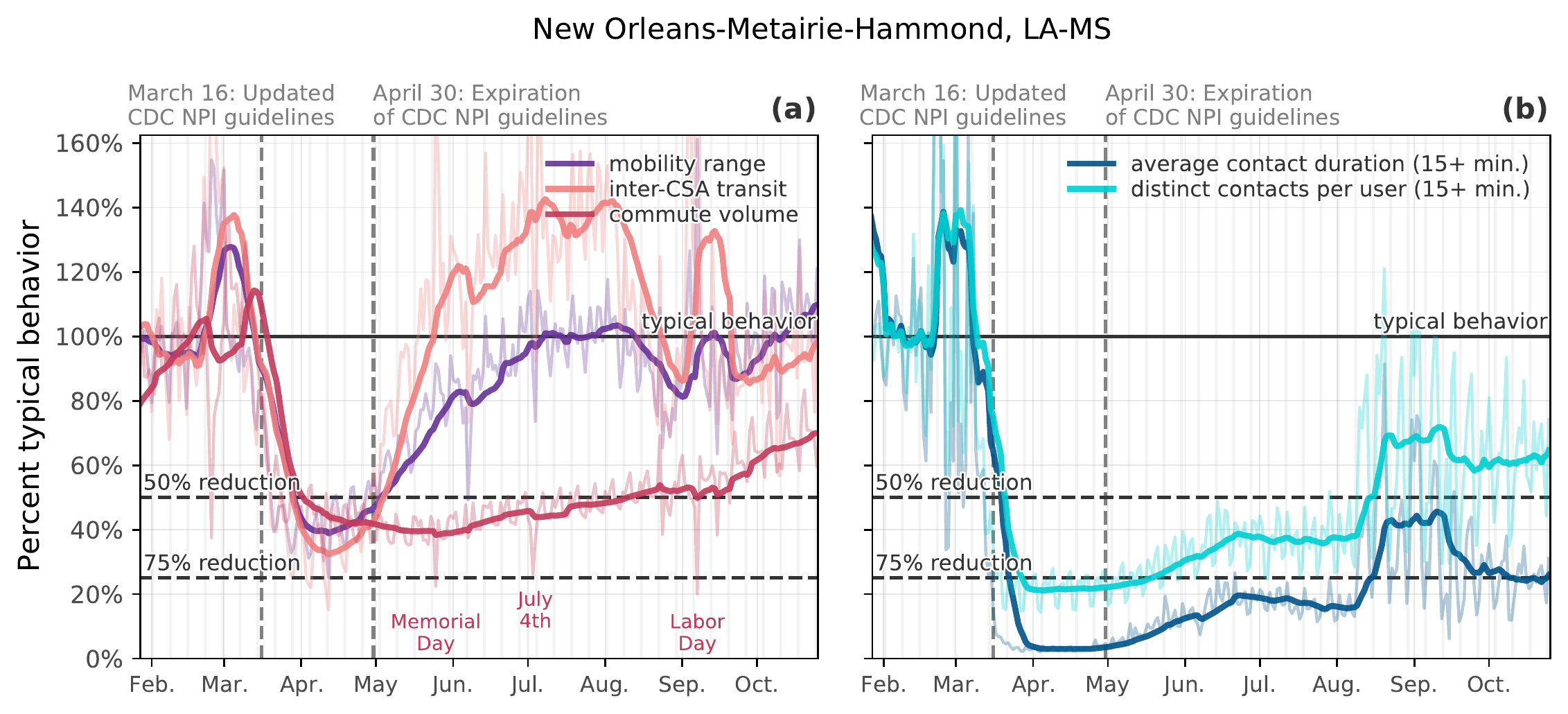}
    \end{subfigure}
    \caption{New Orleans-Metairie-Hammond, LA-MS.}
    \label{fig:city9}
\end{figure}

\begin{figure}[ht]
    \centering
    \begin{subfigure}{0.95\columnwidth}
        \centering
        \includegraphics[width=1.0\columnwidth]{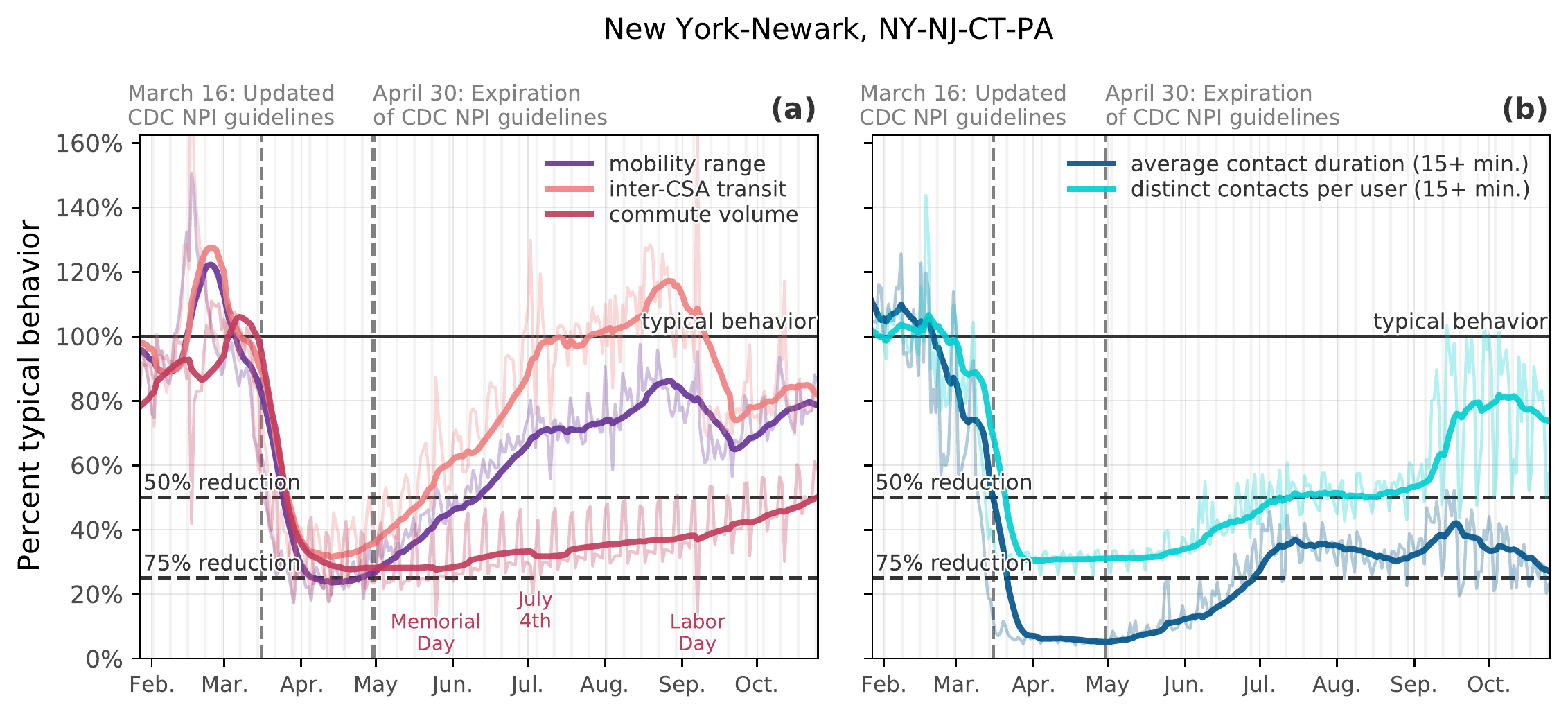}
    \end{subfigure}
    \caption{New York-Newark, NY-NJ-CT-PA.}
    \label{fig:city10}
\end{figure}

\begin{figure}[ht]
    \centering
    \begin{subfigure}{0.95\columnwidth}
        \centering
        \includegraphics[width=1.0\columnwidth]{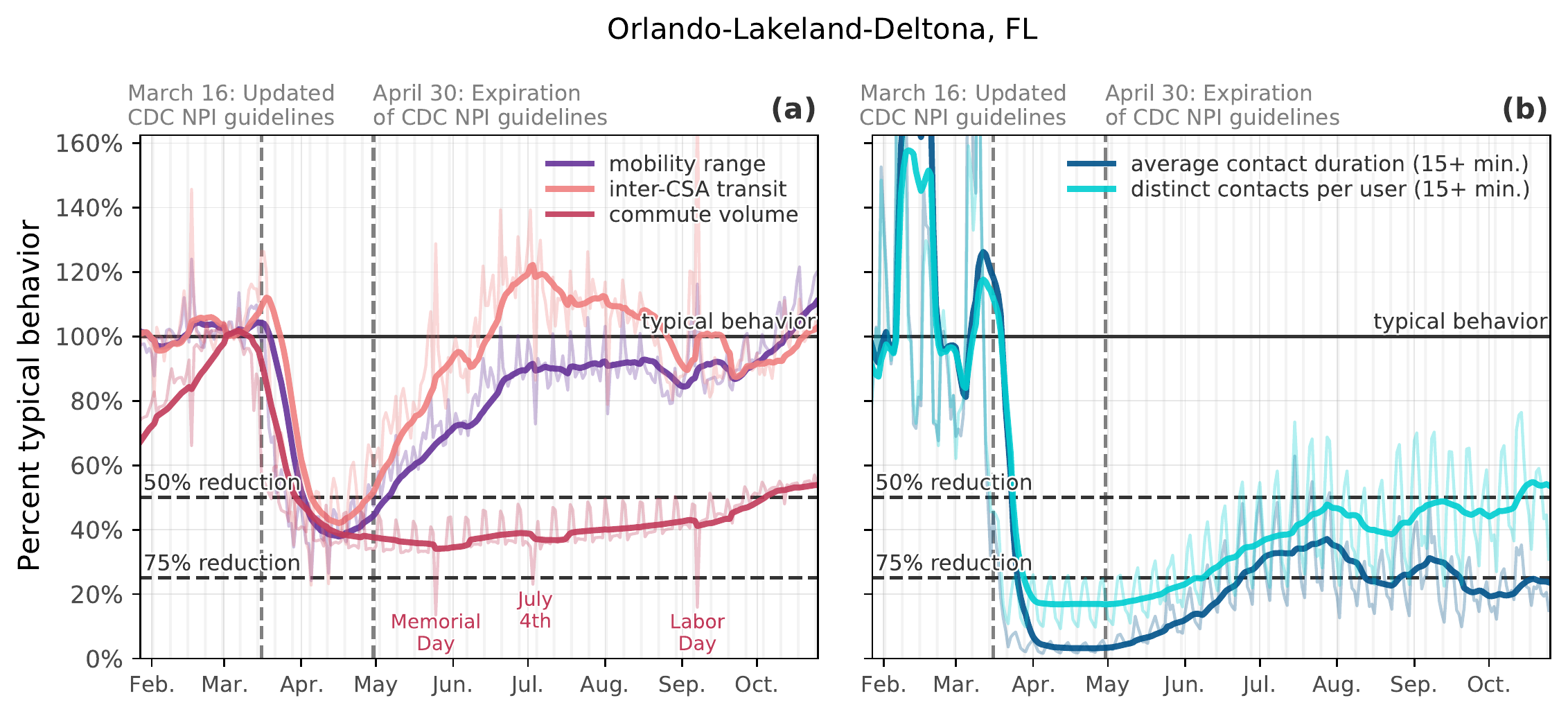}
    \end{subfigure}
    \caption{Orlando-Lakeland-Deltona, FL}
    \label{fig:city11}
\end{figure}

\begin{figure}[ht]
    \centering
    \begin{subfigure}{0.95\columnwidth}
        \centering
        \includegraphics[width=1.0\columnwidth]{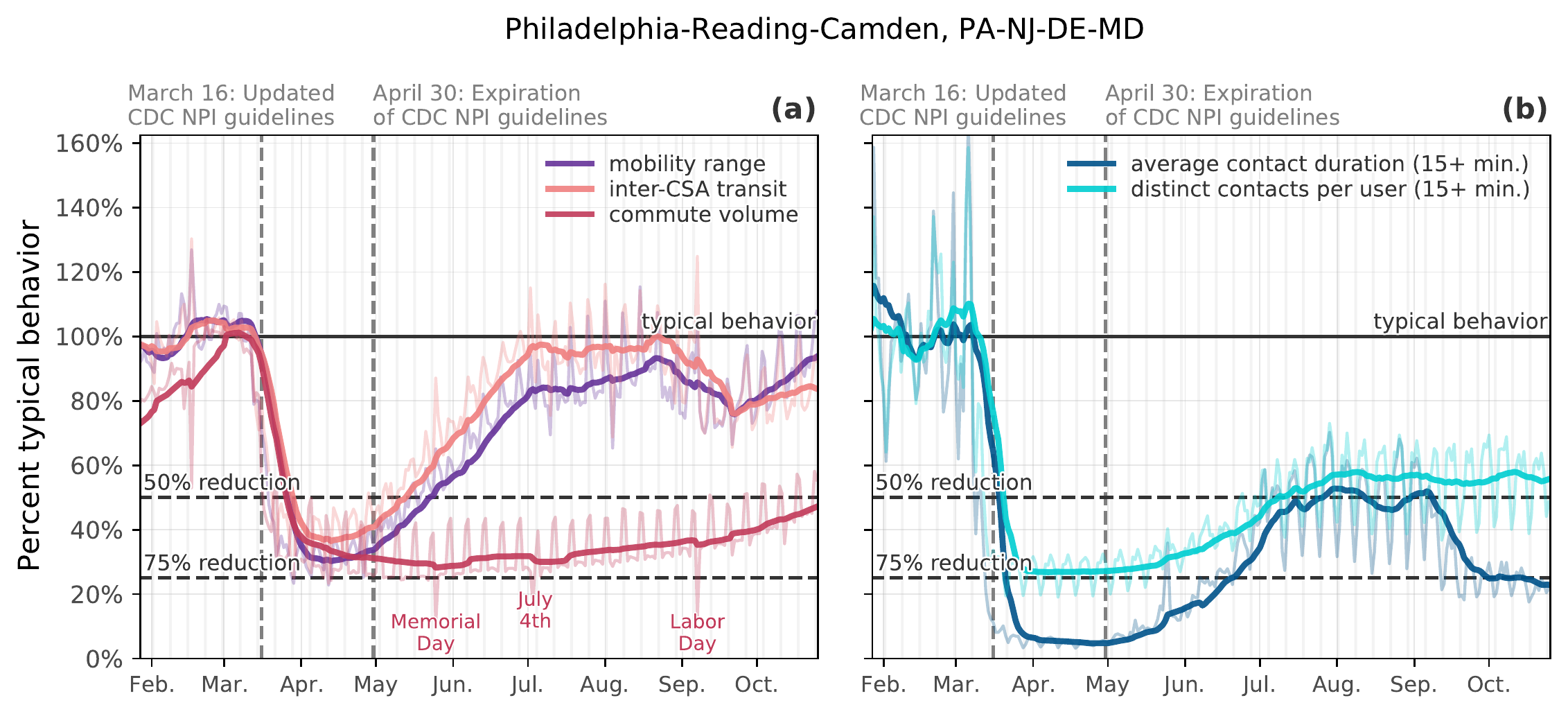}
    \end{subfigure}
    \caption{Philadelphia-Reading-Camden, PA-NJ-DE-MD.}
    \label{fig:city12}
\end{figure}

\begin{figure}[ht]
    \centering
    \begin{subfigure}{0.95\columnwidth}
        \centering
        \includegraphics[width=1.0\columnwidth]{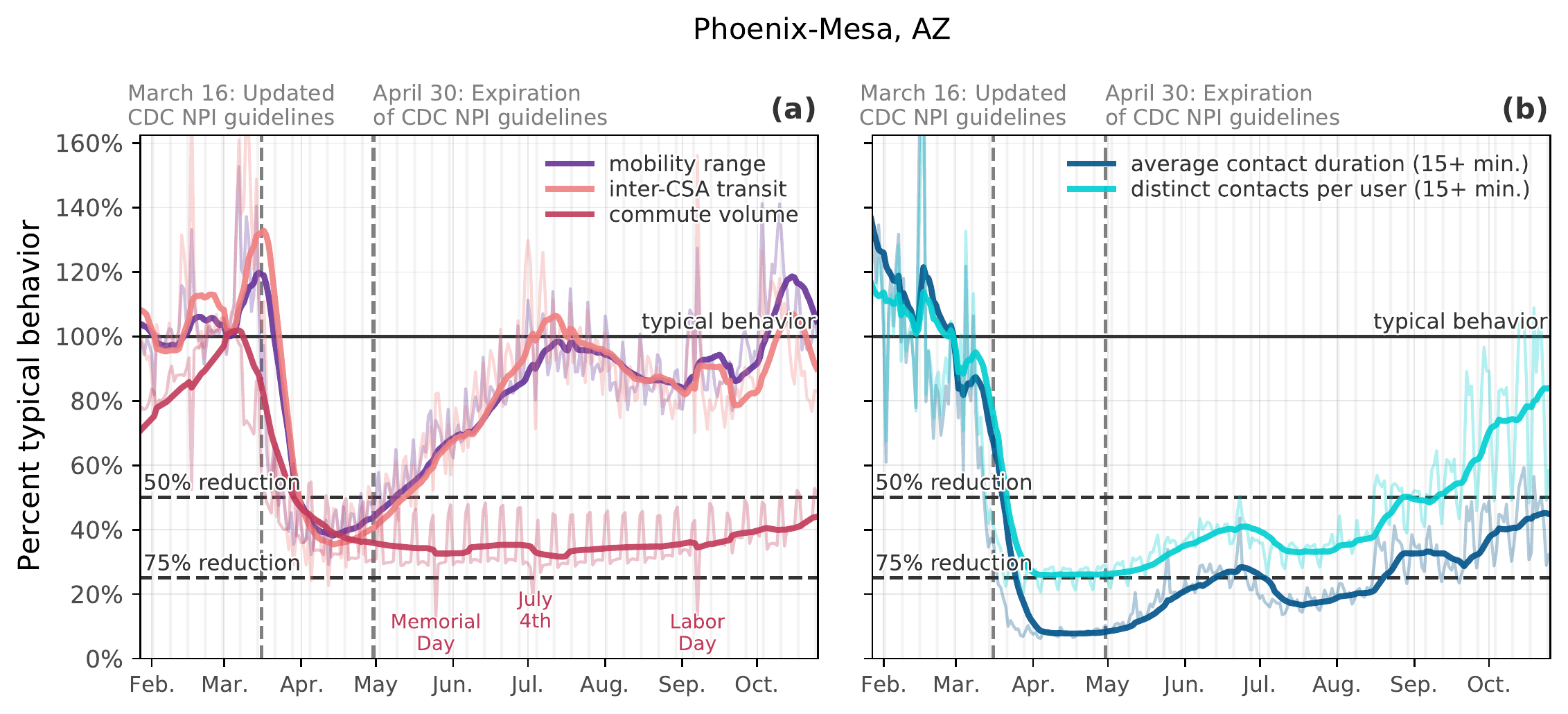}
    \end{subfigure}
    \caption{Phoenix-Mesa, AZ.}
    \label{fig:city13}
\end{figure}

\begin{figure}[ht]
    \centering
    \begin{subfigure}{0.95\columnwidth}
        \centering
        \includegraphics[width=1.0\columnwidth]{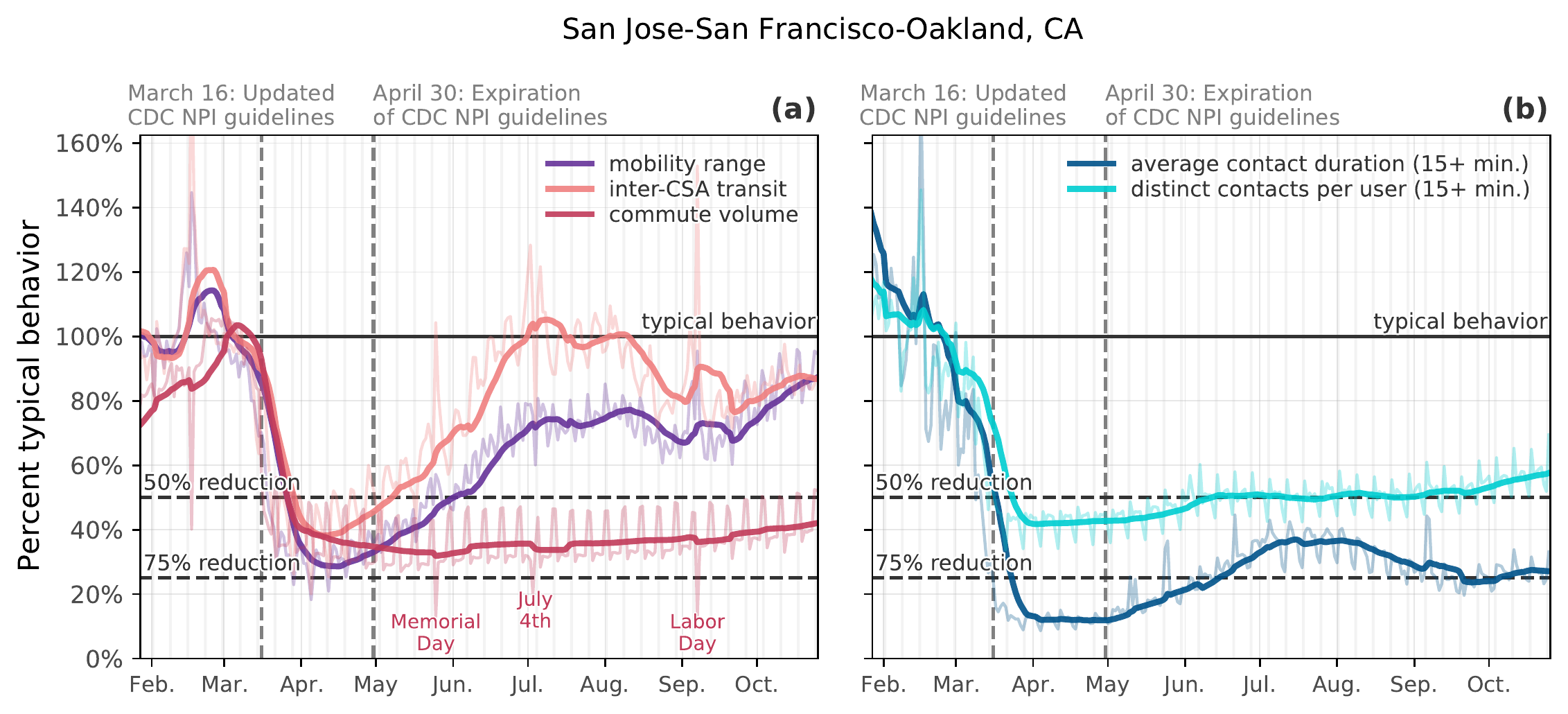}
    \end{subfigure}
    \caption{San Jose-San Francisco-Oakland, CA.}
    \label{fig:city14}
\end{figure}

\begin{figure}[ht]
    \centering
    \begin{subfigure}{0.95\columnwidth}
        \centering
        \includegraphics[width=1.0\columnwidth]{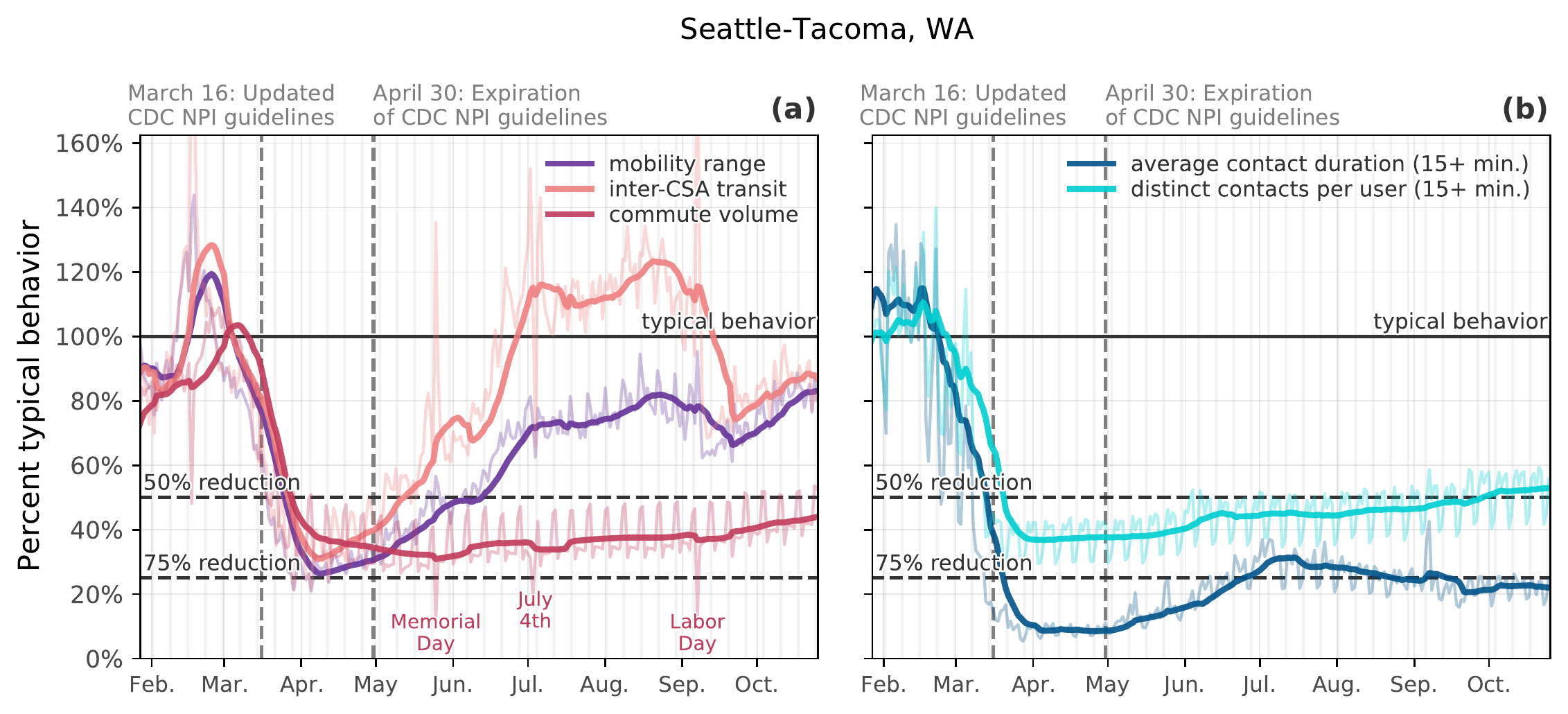}
    \end{subfigure}
    \caption{Seattle-Tacoma, WA.}
    \label{fig:city15}
\end{figure}

\begin{figure}[ht]
    \centering
    \begin{subfigure}{0.95\columnwidth}
        \centering
        \includegraphics[width=1.0\columnwidth]{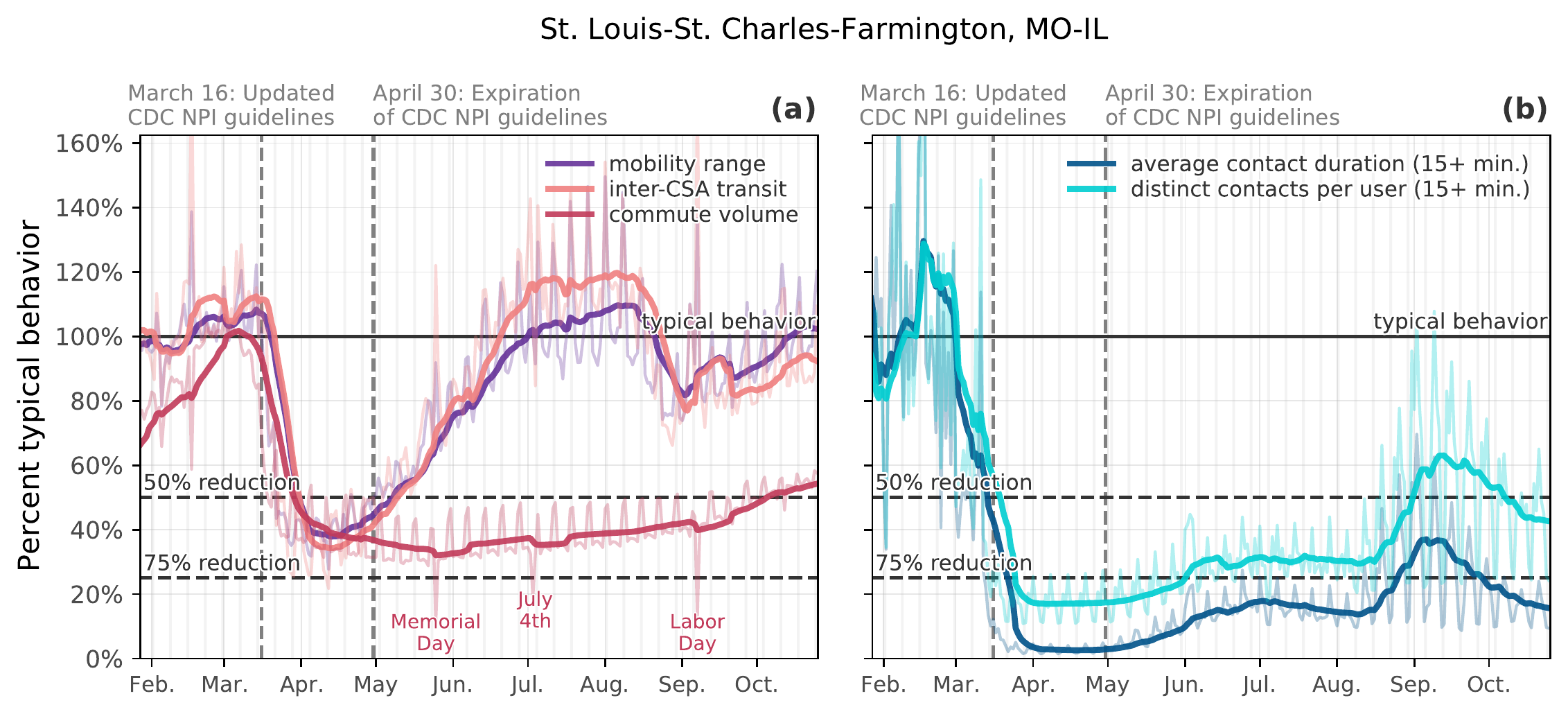}
    \end{subfigure}
    \caption{St. Louis-St. Charles-Farmington, MO-IL.}
    \label{fig:city16}
\end{figure}

\begin{figure}[ht]
    \centering
    \begin{subfigure}{0.95\columnwidth}
        \centering
        \includegraphics[width=1.0\columnwidth]{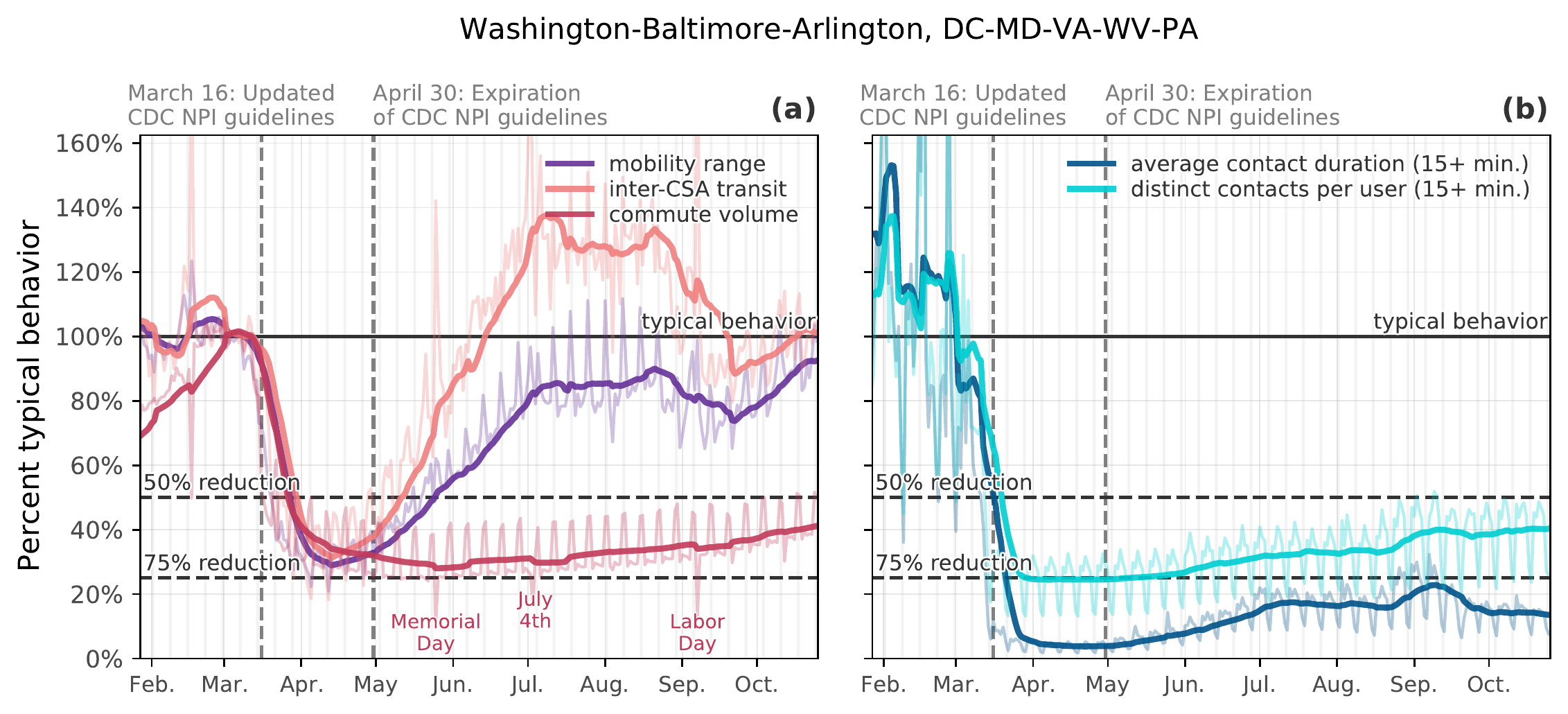}
    \end{subfigure}
    \caption{Washington-Baltimore-Arlington, DC-MD-VA-WV-PA.}
    \label{fig:city17}
\end{figure}

\clearpage
\subsection{Citation diversity statement}\label{sec:citation-div}
Recent work has quantified bias in citation practices across various scientific fields; namely, women and other minority scientists are often cited at a rate that is not proportional to their contributions to the field \cite{Zurn2020, Dworkin2020, Chakravartty2018, Maliniak2013, Dion2018, Caplar2017, Azoulay2020, Ghiasi2018}. In this work, we aim to be proactive about the research we reference in a way that corresponds to the diversity of scholarship in this field. To evaluate gender bias in the references used here, we obtained the gender of the first/last authors of the papers cited here through either 1) the gender pronouns used to refer to them in articles or biographies or 2) if none were available, we used a database of common name-gender combinations across a variety of languages and ethnicities. By this measure (excluding citations to datasets/organizations, citations included in this section, and self-citations to the first/last authors of this manuscript), our references contain 3\% woman(first)-woman(last), 28\% woman-man, 8\% man-woman, 60\% man-man, 0\% nonbinary, 3\% man solo-author, and 0\% woman solo-author. This method is limited in that an author's pronouns may not be consistent across time or environment, and no database of common name-gender pairings is complete or fully accurate.

\printbibliography[title={Supplemental References}]
\end{refsection}
\end{document}